\documentclass[]{aa} % for the letters
\usepackage{graphicx}
\usepackage{subfigure}
\graphicspath{ {images/} }

\usepackage{txfonts}
\usepackage{booktabs}
\usepackage{amsmath}
\usepackage{epstopdf}
\usepackage{enumitem}
\usepackage{multicol}
\usepackage{multirow}
\usepackage{chngcntr}
\usepackage{chngcntr}
\usepackage{soul}
\usepackage{color}
\usepackage{epstopdf}
\usepackage{ulem}
\newcommand{\kms}{km s$^{-1}$\xspace}
\newcommand{\HI}{{\rm H\,{\scriptsize I}}\xspace}
\newcommand{\HII}{{\rm H\,{\scriptsize II}}\xspace}
\usepackage{natbib}
\usepackage[colorlinks, citecolor={blue}]{hyperref}
\begin{document}

\title{Radio properties of the OH megamaser galaxy IIZw 096 }

   \author{Hong Wu
          \inst{1}
          \and
          Zhongzu Wu\inst{1}\fnmsep\thanks{zzwu08@gmail.com}
           \and
          Yu.~Sotnikova,\inst{2}
           \and
         Yongjun Chen\inst{3}
         \and
         Bo Zhang\inst{3}
          \and
         T.~Mufakharov\inst{2,3,5}
          \and
         Zhiqiang Shen\inst{3}
          \and
         Xi Chen\inst{4}
         \and
         A.~Mikhailov\inst{2}
         \and
         M.~Mingaliev\inst{2,5}
         \and
         Xianming L. Han\inst{1,6}
         \and
         Prabhakar Misra\inst{7}
        }

   \institute{College of Physics, Guizhou University, 550025 Guiyang, PR China \email{zzwu08@gmail.com}
              \and
Special Astrophysical Observatory of RAS, Nizhny Arkhyz 369167, Russia
              \and
              Shanghai Astronomical Observatory,
Chinese Academy of Sciences, 80 Nandan Road, Shanghai 200030, PR China \and
Center for Astrophysics, GuangZhou University, Guangzhou 510006, PR China \and
Kazan Federal University, 18 Kremlyovskaya St, Kazan 420008, Russia \and
Dept. of Physics and Astronomy, Butler University, Indianapolis, IN 36208, USA
\and
Department of Physics \& Astronomy, Howard University, Washington, DC 20059,
USA
         }
   \date{  }

% \abstract{}{}{}{}{}
% 5 {} token are mandatory

  \abstract
  % context heading (optional)
  % {} leave it empty if necessaryWe found that continuum emission is resolved with VLBA observations, which indicate that the continuum emission might mainly from starburst emission. The results from two epoch EVN observations of the OH megamaser line emission further%The two components OH maser emission show similar FWHM which is about 70 ~ 99 \kms, they are likely to be two clouds which might relate to the ongoing merging procedures.
   %The HI line profiles at large scale show similar two similar emission features and no clear oribiting velocity field, indicate that the \HI gas might be still at a stage of merging process.
   {
   Based on the two epochs European Very Long Baseline Interferometry (VLBI) Network (EVN) archive data from OH line observations of IIZw 096, we confirm that the high-resolution OH emission in this source mainly comes from two spots (OH1 and OH2) of comp D1 of this merging system. We found no significant variations in the OH 1667 MHz line emission, including flux densities and peak positions. The OH 1665 MHz line emission is detected at about 6 $\sigma$ level in the OH1 region by combining two epoch EVN observations. By using archival data from the Very Long Baseline Array (VLBA), Very Large Array (VLA), and Atacama Large Millimeter Array (ALMA) observations, we investigated the properties of the environment of this component through \HI, CO (3-2), and HCO+ (4-3) lines and the multi-band radio continuum emission. We found that the comp D1 shows the brightest CO, HCO+ line emission, as well as multi-band radio continuum emission. The environment around D1 shows no clear velocity structure associated with circular motions, making it different from most other OH megamasers (OHMs) in the literature, which might have been caused by an effect during the merger stage. Meanwhile, we found that the CO emission shows three velocity structures around D1, including the central broad FWHM region, the double peak region where the CO line profile shows two separated peaks, and the region of the high-velocity clouds where the CO line peaks at a high velocity ($\sim$ 11000 \kms). Similarly, \HI observations in absorption also show high-velocity clouds around the D1 region, which might be due to inflows caused by the merging of two or more galaxy components. Based on the high-resolution K-band VLA and L-band VLBA observations of the radio continuum emission, we derived the brightness temperature in the range $10^{5}$ K to $10^{6}$ K, which is consistent with other starburst dominant OHM sources in the literature. The multi-band VLA observations show that the radio continuum emission of comp D might also have contributions from free-free emission, besides synchrotron emission. As a concenquence, these results support a starburst origin for the OHMs, without the presence of an AGN.
   %It is likely that the dominanated radio continuum emission is from starburst. consequences The CO, HCO+line emission and theK-band VLA-A continuum emission are roughly aligned with abrightest center, The two OH emission regions show an offsetabout 50-75 mas to the central region, and likely tend to the di-rection of double peak region
   %The continuum emission might be uniformly distributed, based on the VLA and VLBA observations.
  }

   \keywords{OH megamaser galaxy: starburst: radio continuum: galaxy radio lines: general.}

  \authorrunning{Wu et al. }            %author_head in even pages
   \titlerunning{radio properties of IIZw 096}  % title_head in odd pages
   \maketitle
%
%________________________________________________ sections below
\section{Introduction}

OH megamasers (OHMs) are luminous 18 cm masers found in (ultra-)luminous infrared galaxies ([U]LIRGs) produced predominantly by major galaxy mergers \citep{2021ApJ...911...38R}. Generally, OH molecules are believed to be pumped by far-IR radiation \citep{2008ApJ...677..985L,2018JApA...39...34H} and triggered by dense molecular gas \citep{2007ApJ...669L...9D}. Both star formation and OHM activity are consequences of tidal density enhancements accompanying galaxy interactions \citep{2007ApJ...669L...9D}. Theoretical studies show that major galaxy mergers are the dominant processes leading to supermassive black hole (SMBH) growth at high masses  ($\geq$$\rm 10^8~M_{\odot}$), which can destabilize large quantities of gas, driving massive inflows towards the nuclear region of galaxies and triggering bursts of star formation \citep{2019NatAs...3...48S}.

The dominant energy source in the central regions of OHM galaxies usually presents features of starburst and active galactic nucleus (AGN), and in turn the radio continuum emission is produced from these activities.  Based on multi-band observations, \cite{2020MNRAS.498.2632H} present a hypothesis that OHM galaxies harbor a recently triggered AGN. High-resolution observations might be essential to determine whether OHM galaxies are hosting an AGN or compact starburst, and their connections with merging stages and other environmental parameters \citep[see][and references therein]{2020A&A...638A..78P}.

IIZw 096 is classified as an LIRG \citep{2010AJ....140...63I}, and is one of the most luminous known OHM galaxies \citep{1986IAUC.4231....2B}. This particular source is the second system to host formal megamasers involving both OH and $\rm H_{2}O$ species \citep{2013A&A...560A..12W,2016ApJ...816...55W}. Optically, IIZw 096 shows complex morphology \citep{2010AJ....140...63I}; it contains four main regions, denoted as A, B, C, D (see Fig. \ref{radio-hst}), where sources A and B are possibly two spiral galaxies. Near-IR imaging and spectroscopic observations show that the source D is a powerful starburst not associated with the primary nuclei (sources A and B), which could be a starburst in the disturbed disk of source A, or even the nucleus of a third galaxy \citep{1997AJ....113.1569G,2010AJ....140...63I}.

  %Although it is believed that this source might contain AGN, whether the radio continuum emission origin AGN is not clear.region D show extremely red colors suggest an older stellar population or an AGN-type nucleus.
 The results from MERLIN observations show that the OH megamaser emission originated from comp. D1 and it is distributed in the form of an elongated structure ($\sim$ 300 pc) with a velocity range of 200 \kms \citep{2011MNRAS.416.1267M}. The estimated lower limit for the enclosed mass is $\sim 10^{9}$ $\rm M_{\odot}$, which is consistent with a massive black hole, and an AGN could also be in this merging system \citep{2011MNRAS.416.1267M}. High-resolution EVN observations \citep{2010PhDT.......280C} found that the OH emission originated from two regions (OH1 and OH2 in Fig. \ref{radio-hst}) and indicated that a new epoch VLBI observations could confirm the assumed structure of the OHM emission and potentially determine the proper motions of the two components.

Generally,  the OH megamasers are produced through significant galaxy merging; however, the environment that facilitates such a phenomenon is still not completely understood, primarily because OHM originating from a central AGN or represents a transition stage between a starburst and AGN  \citep{2018MNRAS.474.5319H}. IIZw 096 is one of the few bright OHM galaxies in OH 1667 MHz line emission, potentially observable for a detailed study of compact megamasers with high-resolution spectral-line VLBI observations. This particular source also contains rich merging components from the infrared and optical observations as noted in the literature \citep[e.g.][]{1997AJ....113.1569G,2010AJ....140...63I,2011MNRAS.416.1267M}. It is likely to be a rare nearby example potentially for studying the environment around the OH megamaser emission regions. The main aims of this paper are to further study the properties of the high-resolution structure of the OHM emission \citep{2010PhDT.......280C}  and find its possible connections with the environmental conditions, including the existence of dense gas, compact radio continuum emission, and also the possible merging status of the pertinent galaxies. The details about radio data collection, reduction, and associated analyses are presented in Section 2. The results and discussion are presented in Sections 3 and 4, respectively. In Section 5, we provide a summary of the primary results and conclusions of this paper.

\section{Data collection, reduction and analysis}
\label{sect:Obs}
\subsection{The archival radio data}
We have collected the archival radio data from EVN, VLBA, VLA, and ALMA. The detailed information about the spectral line projects and supporting multi-frequency continuum data are presented in Table \ref{vlbilist} and Table \ref{vladata}, respectively.
% \st{ The information about the archival data is listed in Table \ref{vlbilist}.
 %To study the arcsecond-scale continuum emission of this source, we also collected the VLA archive data. The %Multi-band VLA projects of this source and the acquired flux densities of comp A and D were summarized in Table %\ref{vladata}.}

 \begin{figure*}
   \centering
\includegraphics[width=15cm]{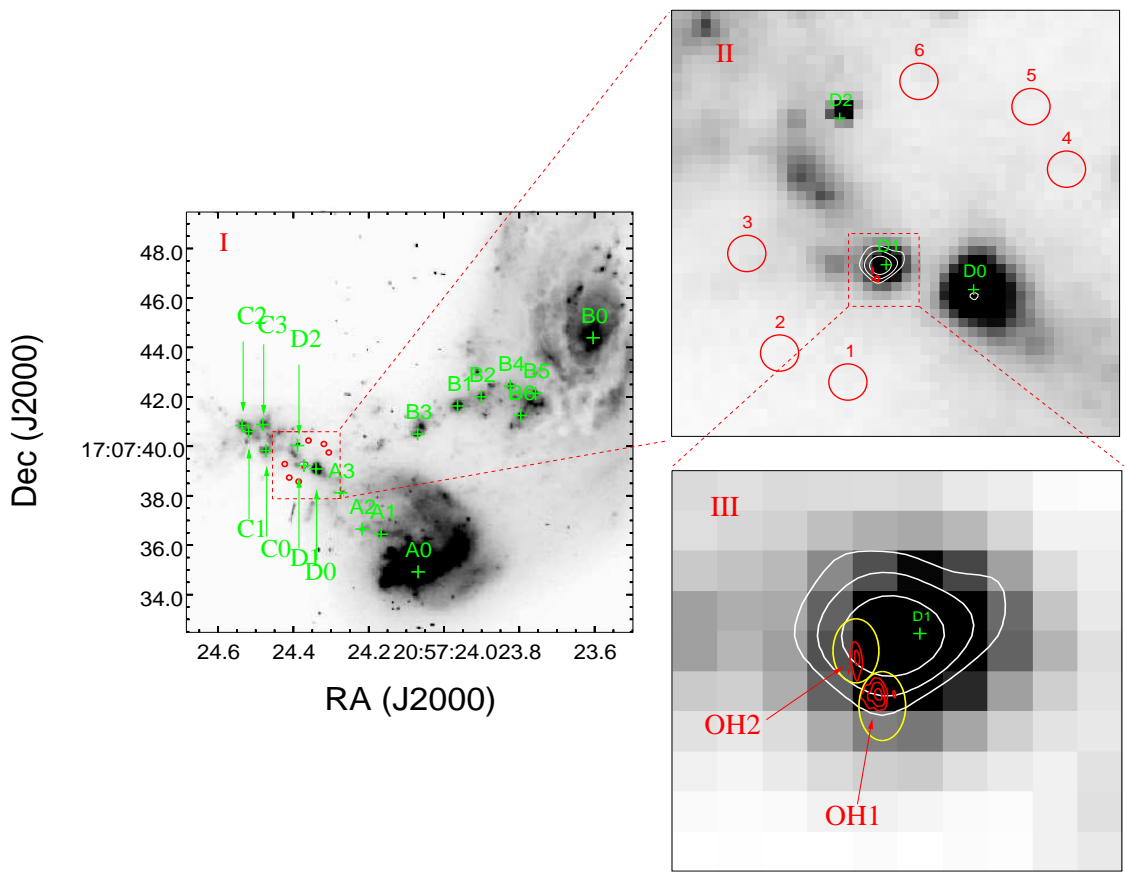}
      \caption{I: The HST-ACS F814W image (grey scale) for IIZw 096. The green crosses indicate the bright spots in this optical image. II: VLA (A configuration) contour map at 33 GHz (white line) are overlaid on the HST image;  the contour levels are 0.0000441 $\times$ (1, 2, 4, 8) Jy/beam and the Beam FWHM: 82.4 $\times$ 59.5 (mas) at -69.1$^{\circ}$. The red circles stand for the regions around the D1 component and the radius is about 0.1 arcsec. The red contour stands for the OH megamaser emission (red) from EVN archival data (project ES064B). The details about image parameters of the OH emission are present in Fig. \ref{EVN2line}. III: The zoomed map of D1 region from II. The yellow ellipses are the two regions where we extracted the integrated OH emission lines.}
    \label{radio-hst}%
\end{figure*}

%%%%%%% ----------------------------------

\begin{figure}
   \centering
  \includegraphics[width=9cm]{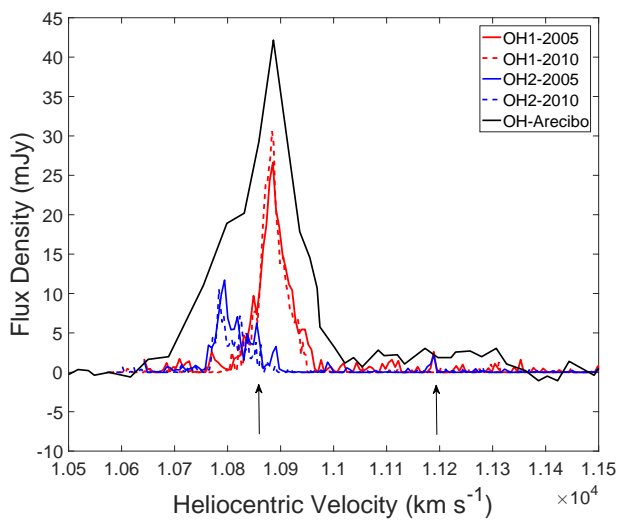}
      \caption{\textbf{The OH line profiles of IIZw 096.}The red and blue spectra are the  OH 1667 MHz line profiles obtained by integrating the signals above 3$\sigma$  over two OH emission regions (OH1 and OH2, see Fig. \ref{radio-hst}) from each channel image of the two EVN line observations as listed in Table \ref{vlbilist}. The black spectrum is the OH profile from Arecibo observations by \cite{1989ApJ...346..680B}. The two arrows represent the velocity of the OH 1667 MHz (left) and 1665 MHz (right) lines based on the optical redshift.}
      \label{evnspectrum}
\end{figure}

%%%%%%-------------------------------------------------
 \begin{figure*}
   \centering
\includegraphics[width=14cm,height=15cm]{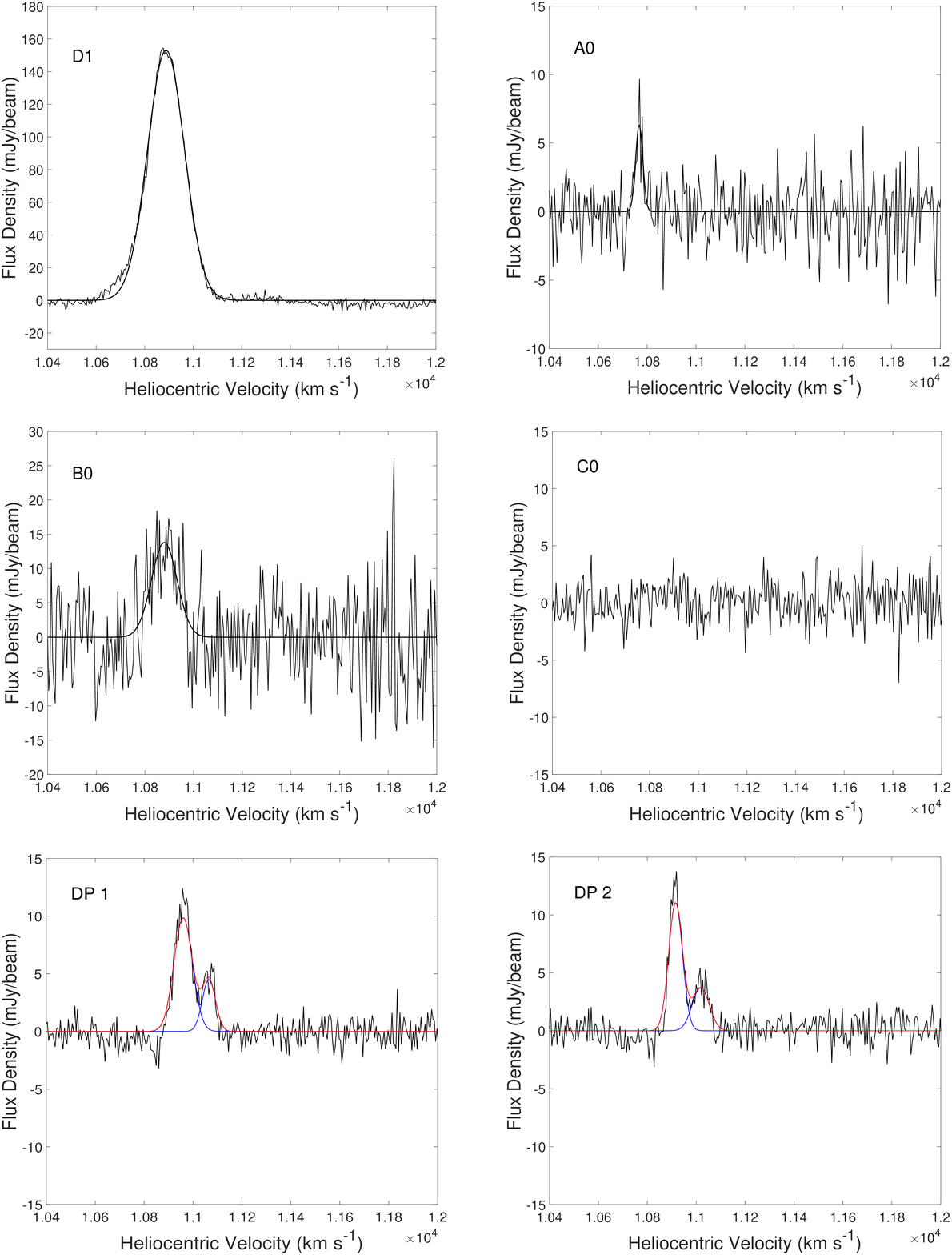}
 \caption{The CO emission lines extracted from regions or components in IIZw 096. The CO line profiles were fitted with one or two Gaussian components.  The blue and red lines are the fitted Gaussian components and the sum of these components, respectively.}
 \label{COabcd}%
\end{figure*}
\begin{figure*}
   \centering

%\includegraphics[width=8cm,height=6cm]{332.eps}
%The two crosses in black color are the OH emission regions OH1 and OH2 as showed in Fig. \ref{radio-hst} and the cross with the corresponding color corresponds to their center coordinate position. The blue line is the HCO+ emission contour averaged channels with velocity ranges from 10763.3-11031.8 km/s,  the contour: 0.00075 $\times$ (1, 2, 4, 8, 16) Jy/beam. The magenta contour is continuum emission from the 33 GHz VLA-A observation with contour level: 0.0000441 $\times$ (1, 2, 4, 8) Jy/beam. The green contour is the CO line emission at the channel with peak velocity about 10887 km/s with contour: 0.0063 $\times$ (1, 2, 4, 8, 16) Jy/beam.
%\includegraphics[width=8cm,height=6cm]{222.eps}
\includegraphics[width=16cm,height=6cm]{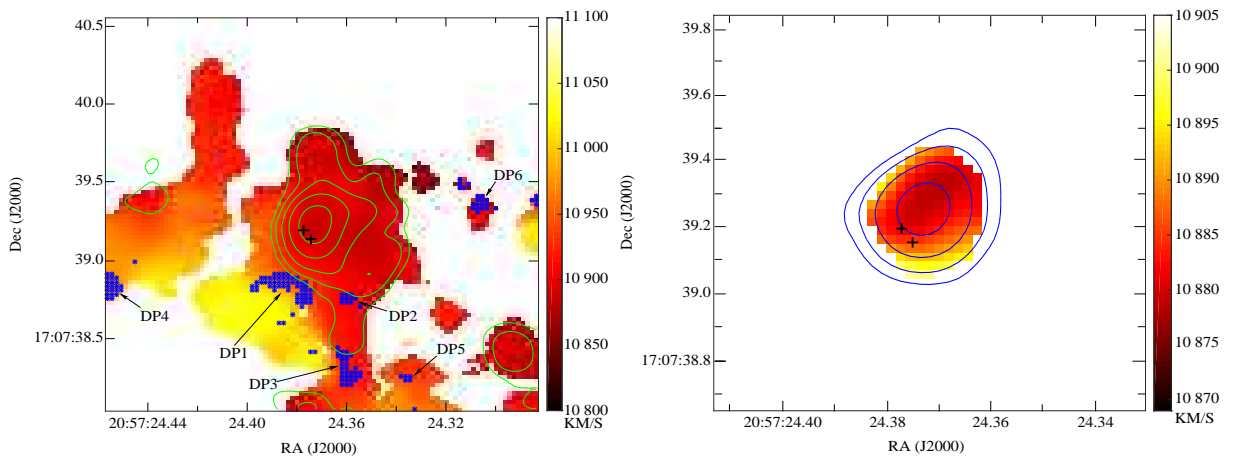}
      \caption{The velocity structure of CO (left panel) and HCO+ (right panel) line emission ($>$10 mJy/beam) around D1 region. The two crosses in black color are the OH emission regions OH1 and OH2\textbf{,} as shown in Fig. \ref{radio-hst}. The green and blue contour stand for the CO and HCO+ emissions, respectively. The contour levels are present in the caption of Fig. \ref{cohcooh33}. The blue color spots stand for where the extracted spectrum shows separated double peaks, roughly distributed in 6 regions (DP1 - DP6). }
    \label{co-hco1115}%
\end{figure*}
 \begin{figure*}
   \centering
 \includegraphics[width=16cm,height=6cm]{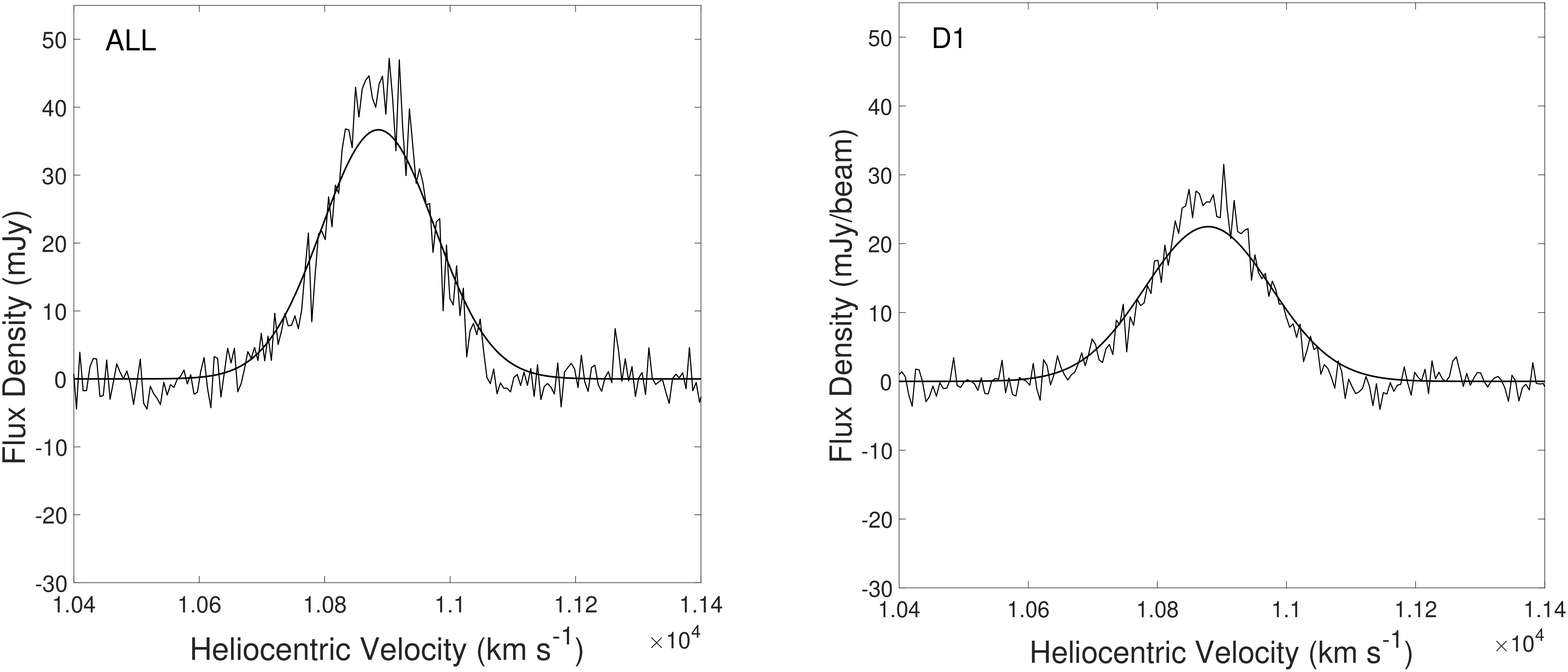}
 \caption{\textbf{The HCO+ lines in IIZw 096.}Left panel: The integrated HCO+ spectrum extracted in a region with a size of 0.45' $\times$ 0.45' centered at D1. Right panel: the HCO+ emission line emission extracted at D1 spot. The line profiles were fitted with one Gaussian component.}
 \label{HCOabcd}%
\end{figure*}

%--------------------------------------------------------------------
\begin{figure*}
   \centering
   \includegraphics[width=14cm,height=5.5cm]{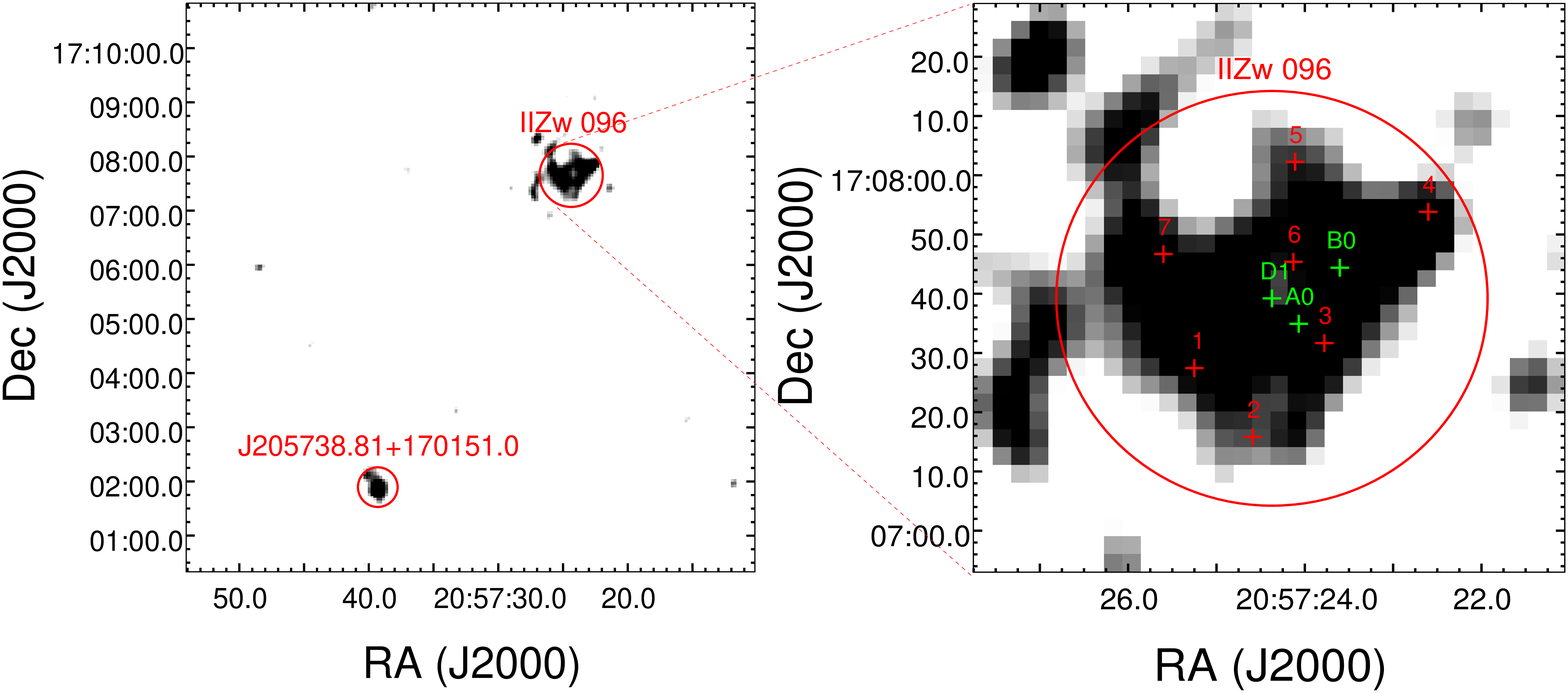}

      \caption{The \HI emission channel image (V $\sim$ 10849.3 km/s) of IIZw 096. Left panel: The \HI emission of IIZw 096 is mainly distributed in a region with size of 70" $\times$ 70" centered at D1 for IIZw 096; The \HI emission for the new galaxy is centered at RA: 20 57 39.307, Dec: +17 01 53.762 with a region size of 44" $\times$ 44". The right panel: The zoomed image of IIZw 096, the plus sign stands for the spots where we extracted the \HI spectrum.}
      \label{vlaHI}

\end{figure*}
%--------------------------------------------------------------------

  %-------------------------------------
   \setlength{\tabcolsep}{0.05in}
  \begin{table*}
       \caption{Parameters of the high resolution spectral line observations. }
     \label{vlbilist}
  \centering
  \begin{tabular}{c c c c l c c c c c }     % 8 columns
  \hline\hline
   Observing Date & Frequencies & Line& Array                                             & Phase       & Program    & $\Delta_V$    & Beam  &  P.A.   & rms \\
                  &    (GHz)  &  &                                      & Calibrator  &            &  \kms        & (") $\times$(")   &  ($\circ$) & (mJy/beam)   \\
   \hline
     2005Jun08    &  1.6    &  OH  & $\rm EVN^{1}$   & J2052+1619$\rm ^{pr}$  & EK020      & 6.1               & 0.005 $\times$0.008 & -60 &0.26\\
     2010Jun08    &  1.6    &  OH  & $\rm EVN^{2}$               & J2052+1619$\rm ^{pr}$  & ES064B     & 3.1               & 0.028$\times$0.005 & 9  &0.67 \\
    % 2014May18/19 &  1.5   & Cont.    & VLBA                                              & %J2052+1619$\rm ^{pr}$  & BS0233     &         -                & \textbf{0.008 $\times$0.003} &\textbf{ -2 }&\textbf{0.015} \\
     %2014Apr18    &  33    & Cont.    & VLA-A                             & J2139+1423  & 14A-741    & \textbf{-}               & \textbf{0.082$\times$0.070}&\textbf{ -69}  &\textbf{0.015}\\

     2015Jul16    &  333.3 & CO (3-2)  & ALMA  & J2051+1743 & 2012.1.01022.S  &  5.4      & 0.20$\times$0.17  &-54& 1.91    \\
     2015Jul16    & 343.9 &HCO+ (4-3) & ALMA  & J2051+1743 & 2012.1.01022.S  &  5.4      & 0.19$\times$0.16  &-55 & 1.92   \\
     2001Jul22    &  1.4   & \HI      & VLA-C                     & 2035+189& AG0613 & 22.2          & 19.7$\times$17.9  &72& 0.33 \\
   \hline
   \end{tabular}
   \vskip 0.1 true cm \noindent Notes. Column (3): The atom or molecular lines. Column (4): The observational array. $\rm EVN^{1}$: Ef, Jb, CM, Nt, Tr, Wb, On, Mc, Ur, Hh, Ar.  $\rm EVN^{2}$: Ef, Jb, On, Mc, Tr, Sh, Ur, Wb. Column(5): The phase-reference calibrator for EVN projects and phase calibrator for VLA and ALMA projects. The index 'pr' means the project is in phase-reference mode. Column (6): The program name. Column (7): The spectral velocity resolution in \kms. Column (8) and (9): The beam FWHM and position angle. Column (10): The 1 $\sigma$ noise level for channel image.
   \end{table*}

%--------------------------------------------------------------------
%--------------------------------------------------------------------
\begin{table*}
\small
\caption{The parameters of the emission lines detected in IIZw 096}
\label{allline}
\begin{tabular}{lllllll}

\hline\hline
Line & components & RA(J2000)  Dec(J2000)       & Gauss amplitude   & Gauss center      & Gauss FWHM       & Gaussian area       \\
 &            & (hh mm ss,+dd mm ss)        & mJy               & \kms              & \kms             & mJy*\kms       \\
  \hline
 %2010-OH    & 20 57 24.375, 17 07 39.146  &  17.9$\pm$4.1     & 10883.0$\pm$12.4  & 135.2$\pm$24.8    & 2174.98$\pm$707.87     \\
OH1667
 & 2005-OH1   & 205724.375+170739.144  &  21 $\pm$ 3     & 10887 $\pm$ 1   & 67 $\pm$ 2     & 1429 $\pm$ 220      \\
 & 2005-OH2   & 205724.377+170739.186  &  7.2 $\pm$ 1.1     & 10805 $\pm$ 2  & 82 $\pm$ 3    & 588 $\pm$ 94     \\
 & 2010-OH1   & 205724.375+170739.145  &  25 $\pm$ 4     & 10884 $\pm$ 1   & 53 $\pm$ 1    & 1359 $\pm$ 198     \\
 & 2010-OH2   & 205724.377+170739.187  &  5.5 $\pm$ 0.8     & 10803 $\pm$ 1  & 74 $\pm$ 2    & 414 $\pm$ 63     \\
  \hline
CO
& D1:          & 205724.372+170739.221  & 153 $\pm$ 1$^{b}$    & 10888 $\pm$ 1    & 179 $\pm$ 1    & 29176 $\pm$ 155$^{b}$   \\
&A0:           & 205724.069+170734.921  & 6.4 $\pm$ 1.2$^{b}$    & 10768 $\pm$ 3    & 34 $\pm$ 7    & 232 $\pm$ 46$^{b}$    \\
&B0:           & 205723.604+170744.387  & 14 $\pm$ 2$^{b}$    & 10880 $\pm$ 8    & 124 $\pm$ 19   & 1821 $\pm$ 257$^{b}$   \\
&C0:           & 205724.473+170739.826  & --                 & --                   & --                 & --                   \\
& DP1       &        & 10 $\pm$ 2         & 10960 $\pm$ 7    & 89 $\pm$ 15      & 634 $\pm$ 217     \\   % 1
 &     &                & 4.5 $\pm$ 1.3         & 11065 $\pm$ 14   & 63 $\pm$ 27      & 303 $\pm$ 146     \\   % 2
%   &   &                & 9.9$\pm$1.8         & 10959.7$\pm$11.6   & 92.0$\pm$23.2      & 1236.9$\pm$282.9     \\   % 1+2
\hline
HCO+
&ALL           & 205724.372+170739.221  & 43 $\pm$ 1    & 10886 $\pm$ 1    & 180 $\pm$ 3   & 8283.5 $\pm$ 151.9     \\
&D1            & 205724.372+170739.221  & 27 $\pm$ 1$^{b}$    & 10880 $\pm$ 2    & 196 $\pm$ 4   & 5509 $\pm$ 102$^{b}$     \\
\hline
\HI
&  total      & 205724.372+170739.221   & 12$\pm$ 1      & 10799 $\pm$ 5     & 183 $\pm$ 12      & 2391 $\pm$ 143   \\%70x70(以D1为圆心画一个35为半径的圆)
&             & --                      & -2.9 $\pm$ 0.4      & 11016 $\pm$ 38    & 53 $\pm$ 91      &- \\ %-164.89 $\pm$262.18  \\

%&             &                         &                    &                       &                      &                   \\
&D1           & --                      &  1.8 $\pm$ 0.5$^{b}$       & 10780 $\pm$ 17      & 135 $\pm$35      & 247 $\pm$ 90$^{b}$  \\
&             & --                      &  1.0 $\pm$ 0.4$^{b}$       & 10863 $\pm$ 19      & 49  $\pm$38      & 54 $\pm$ 44$^{b}$  \\
&             & --                      & -1.6 $\pm$ 0.6$^{b}$       & 10994 $\pm$ 18      & 91  $\pm$42      &-\\ %-154.14$\pm$65.79$^{b}$  \\%xishou
 %44X44     %44X44
&J205738.81+170151.0 %  & 205739.307+170153.762   &  4.4$\pm$1.0         & 10817.8$\pm$12.7   & 224.8 $\pm$25.4     & 889.8$\pm$241.5   \\ %44X44
                     & 205739.307+170153.762                           &  2.9 $\pm$ 0.8         & 10689 $\pm$ 18   & 108 $\pm$ 35     & 341 $\pm$ 136   \\
           &          & --                           &  4.4 $\pm$ 0.9         & 10820 $\pm$ 12   & 118 $\pm$ 25     & 549 $\pm$ 155   \\
 \hline
%$H_{2}$O & 1+2        & 205723.9+170739.0     &  4.6$\pm$1.0      & 10790.1$\pm$15.8  & 250.0$\pm$31.7    & 1048.8$\pm$296.7     \\
% & 1      & --                          &  4.6$\pm$1.0      & 10787.3$\pm$14.3  & 141.4$\pm$28.7    & 687.2 $\pm$195.2     \\
% & 2      & --                          &  2.5$\pm$0.8      & 10942.1$\pm$25.7  & 135.9$\pm$51.3    & 361.6 $\pm$172.5     \\

\hline
\end{tabular}
\vskip 0.1 true cm \noindent Columns (1): The line type; (2): The emission components; (3): RA and Dec of the components or spots; Columns (4)-(7) The Gaussian fitted model parameters, where 'b' stands for the units for those values that should be divided by beam.
\end{table*}

%-------------------------------------------------------------------

\subsection{Data reduction}
The VLBI data (EVN and VLBA) were calibrated using the NRAO Astronomical Image Processing System (AIPS) package. The main procedures of the VLBI data reduction include ionospheric correction, amplitude calibration, editing, bandpass calibration, instrumental phase corrections, antenna-based fringe-fitting of the phase calibrator, and subsequently applying the solutions to the target source.
The EVLA data were calibrated using the pipeline of the Common Astronomy Software Application package \citep[CASA][]{2007ASPC..376..127M}. The calibration of historical VLA data was done in AIPS following standard procedures. For determining accurately the velocities of the line emission, we have also corrected for the effects of the Earth's rotation and its motion within the Solar System.

We imported all the calibrated data into the DIFMAP package \citep{1997ASPC..125...77S} to obtain the continuum and spectral line (OH and \HI) channel images. The EVN data at epoch 2005Jun08 (see Table 1) show a high signal-to-noise ratio for the peak line channel image (S/N $>$ 60). We have cleaned the brightest region in this peak line channel image (V $\sim$ 10886 \kms) and use these clean models to calibrate the data (phase-only self-calibration) prior to making all the channel images, which is similar to the self-calibration procedure performed by \cite{2010PhDT.......280C} for this project. We conduct no self-calibration for other continuum and line emission projects. The multi-band VLA observations have different resolutions and show extended structures (see Fig. \ref{L-Ka1}). We have remade these images with the same cell size and restored the beam to be 2" $\times$ 2" \citep[the beam size of the VLA-A L-band map, see][]{2015A&A...574A...4V}. Subsequently, we measured the total flux densities and uncertainties of comp D and A (see Table \ref{vladata}) by fitting with a single component using task "imfit" in the CASA package.

\begin{figure*}
   \centering
   \includegraphics[width=16cm,height=6cm]{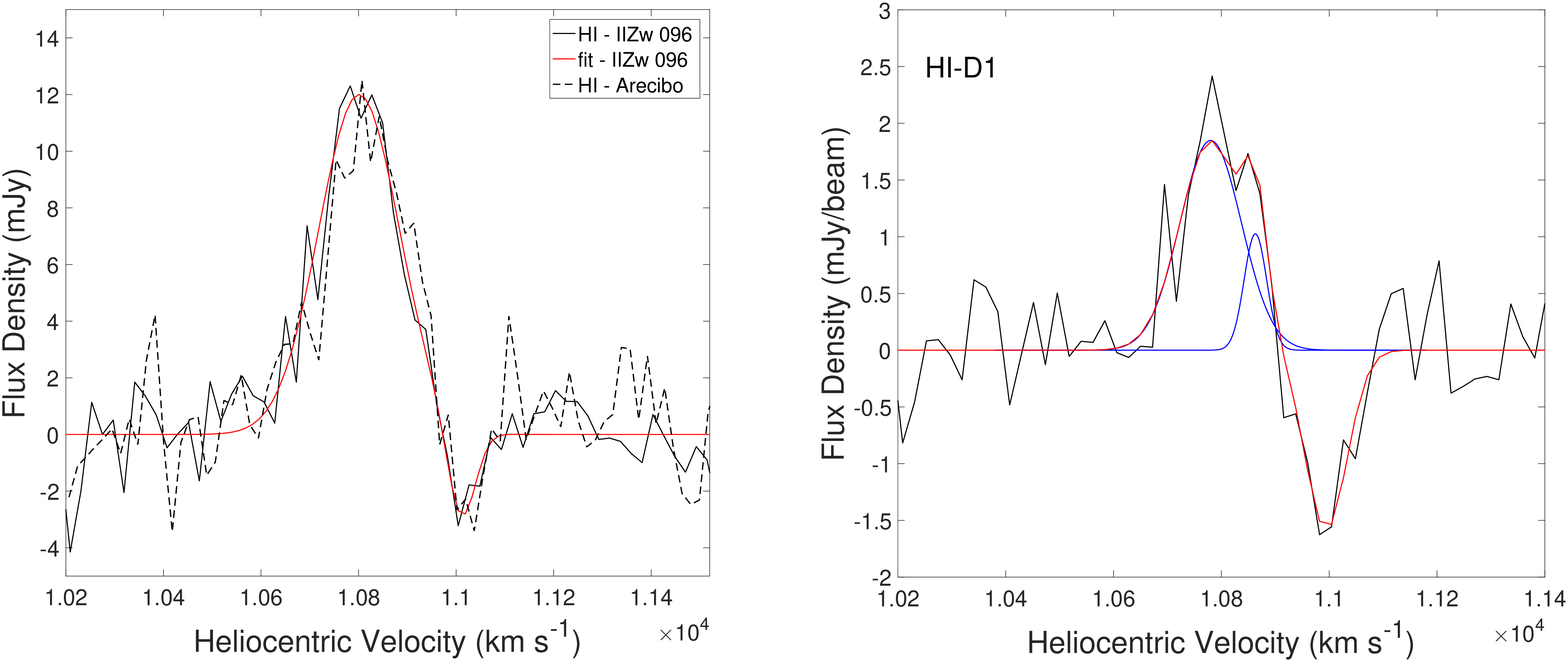}
      \caption{The \HI emission profiles of IIZw 096. The black solid line profile is the detected \HI spectrum, the blue and red lines are the fitted Gaussian components and the sum of those components, respectively. The dashed line from the left panel is the \HI line profile from Arecibo observations made by \cite{2015MNRAS.447.1531C}.
      }
      \label{HItotal}
\end{figure*}

%--------------------------------------------------------------------

\begin{figure*}
   \centering
 \includegraphics[width=14cm,height=5.5cm]{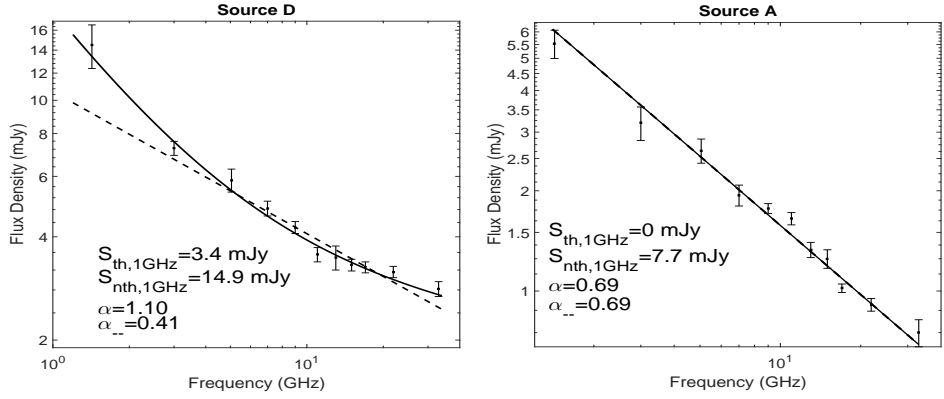}
      \caption{\textbf{The radio spectral index by integral flux of IIZw 096.} Radio continuum spectra of integrated flux densities of D and A from multi-band VLA projects listed in Table \ref{vladata} . The dashed and solid lines stand for fitting results from  equations: a$\times$$\nu$$^{-\alpha}$ and $S_{th} $$\times$ $\nu$$^{-0.1}$+$S_{nth}$ $\times$ $\nu$$^{-\alpha}$, respectively; where $S_{th} $ and $S_{nth}$ stands for non-thermal (synchrotron) and thermal (free-free) flux densities, respectively. }

    \label{ConstantABD}%
\end{figure*}
 \begin{figure}
   \centering
\includegraphics[width=8cm]{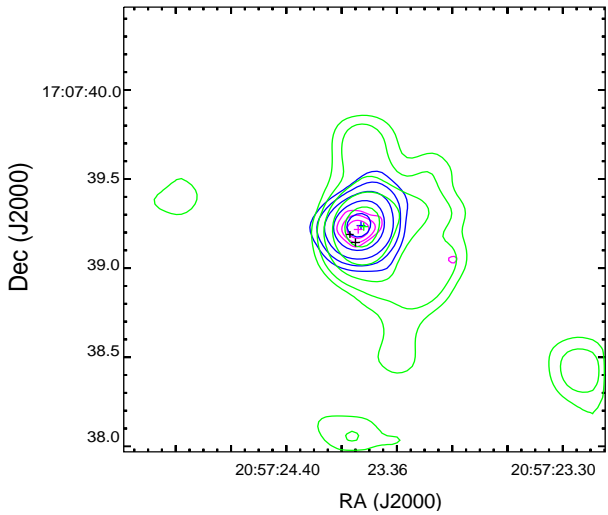}
      \caption{\textbf{The radio contour levers map of IIZw 096.}The two crosses in black color are the OH emission regions OH1 and OH2 as shown in Fig. \ref{radio-hst}\textbf{,} and the cross with the corresponding color corresponds to their center coordinate positions. The blue line is the HCO+ emission contour averaged channels with velocity in the range 10763.3-11031.8 km/s, with the contour level: 0.00075 $\times$ (1, 2, 4, 8, 16) Jy/beam. The magenta contour represents the continuum emission from the 33 GHz VLA-A observation with contour level: 0.0000441 $\times$ (1, 2, 4, 8) Jy/beam. The green contour is the CO line emission at the channel with a peak velocity about 10887 km/s, with contour: 0.0063 $\times$ (1, 2, 4, 8, 16) Jy/beam.}
    \label{cohcooh33}%
\end{figure}
%--------------------------------------------------------------------
\begin{figure*}
   \centering
\includegraphics[width=18cm]{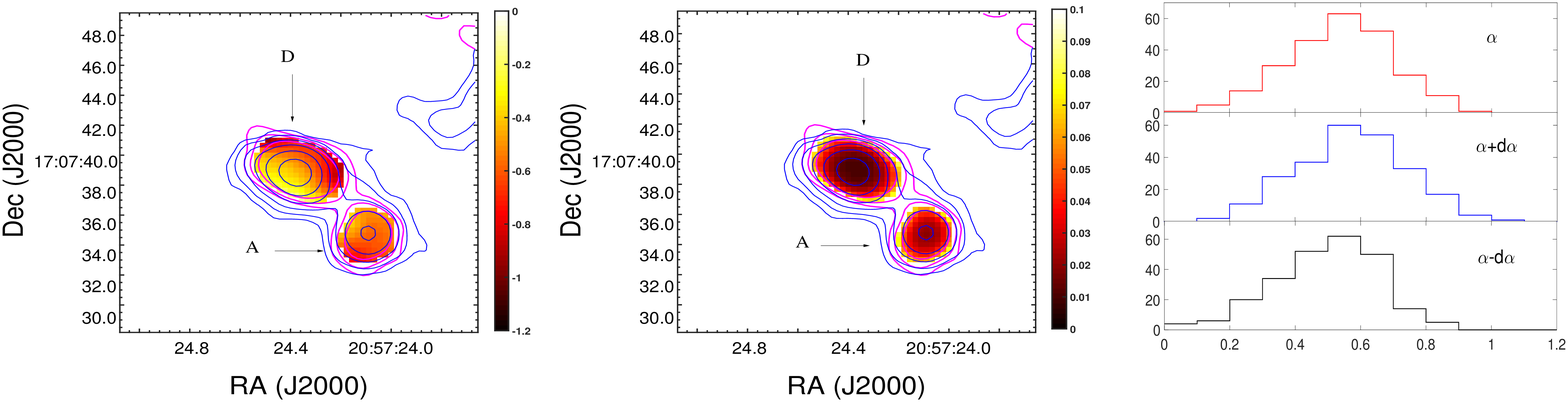}
\includegraphics[width=18cm]{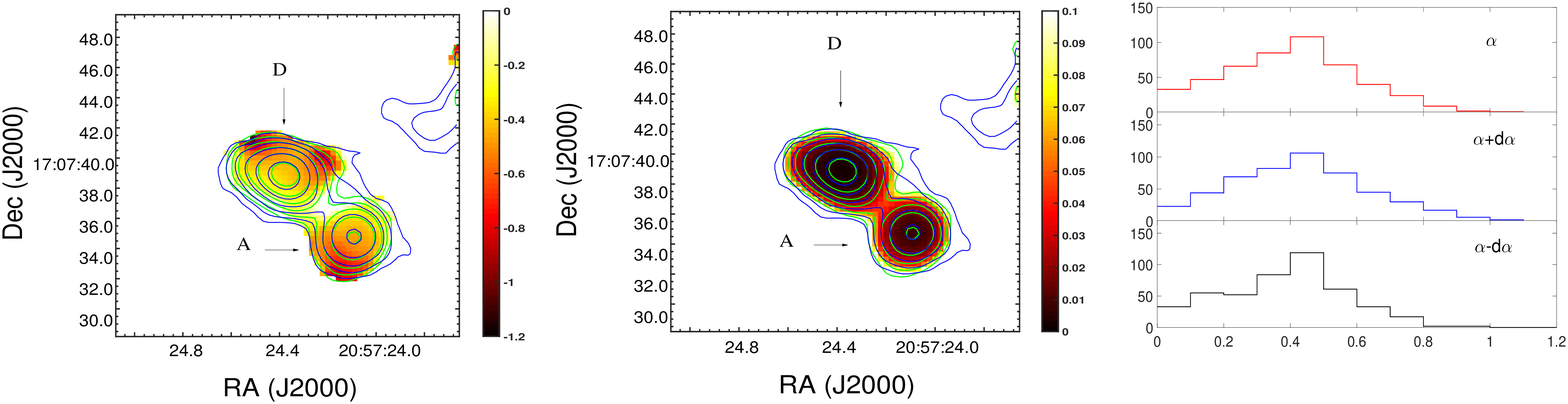}
\includegraphics[width=18cm]{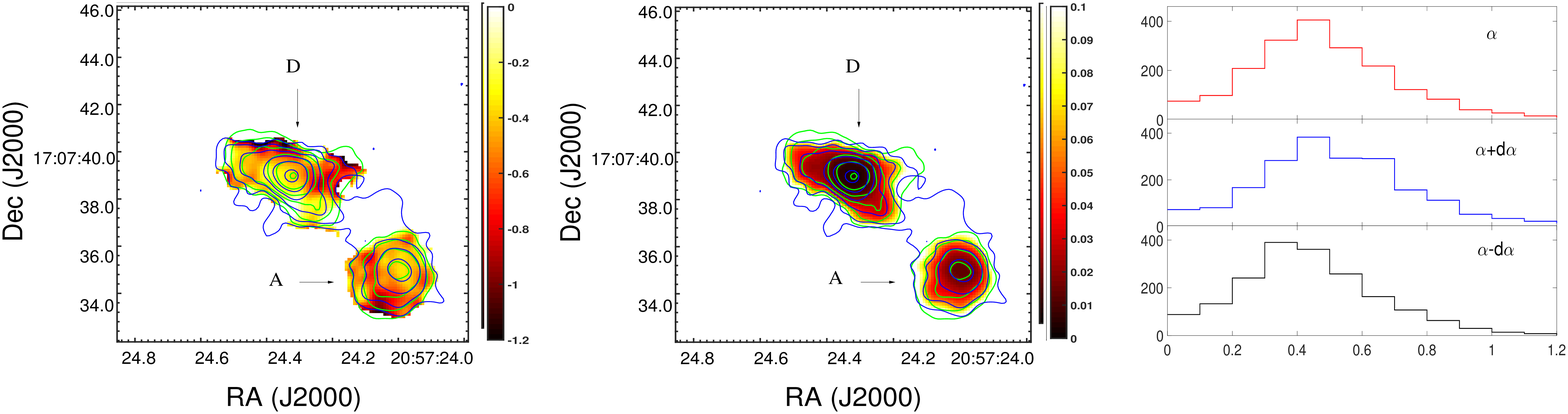}

      \caption{\textbf{The resulting $\alpha-$maps of IIZw 096.} The color maps of each row stand for the radio-spectral-index ($\alpha$, left panel) and its uncertainties $d_{\alpha}$ (middle panel) derived from each pixel value, the colour bar shows the $\alpha$ values from low to high present in the right of each figure. The right panel presents the histograms of the $\alpha$-maps: the $\alpha$, $\alpha$+$d_{\alpha}$ and $\alpha$-$d_{\alpha}$ (from top to bottom). The uncertainties of $\alpha$ were derived from following equation: $(d_{\alpha}=-\frac{\sqrt{(0.4343*rms_{1}/S_{{\upsilon}_{1}})^2+(0.4343*rms_{2}/S_{{\upsilon}_{2}})^2}}{log(\upsilon_{1}/\upsilon_{1})})$, where the rms corresponds to the noise of the radio image, and the $S_{\upsilon}$ is the flux densities at each pixel. The top row is the result obtained from images at 1.4 GHz and 9 GHz, with restored beam: 2" x 2". The middle row is the result obtained from images at 3 GHz and 9 GHz, with restored beam: 2" $\times$ 2". The bottom row is obtained from images at 3 GHz and 9 GHz, with restored beam: 1" $\times$ 1". The contour levels: magenta (1.4 GHz): 0.000335 $\times$ 1, 2, 4 mJy/beam,  blue (9 GHz) : 0.000069 $\times$ 1, 2, 4, 8, 16, 32 mJy/beam.  green (3.0 GHz): 0.000119 $\times$ 1, 2, 4, 8, 16, 32 mJy/beam. }
    \label{359}%
\end{figure*}
%--------------------------------------------------------------------
%--------------------------------------------------------------------

%--------------------------------------------------------------------

\section{Results}
\subsection{Two epoch EVN results of OH megamaser emission}
  \cite{2010PhDT.......280C} has presented the results from one-epoch EVN observation (project EK020, see Table \ref{vlbilist}) and found that the high-resolution OH emission is distributed in two regions as OH1 and OH2 from the D1 component (see Fig. \ref{radio-hst}). Based on the two epoch spectral-line EVN observations (EK020 and ES064, see Table \ref{vlbilist}), we have investigated the integrated OH 1667 line spectrum by averaging the 3 $\sigma$ signals from each channel image over two elliptical regions for OH1 and OH2 (see Fig. \ref{evnspectrum}). We can see that the detected OH line profiles are partly resolved and consistent with the Arecibo observations.
  %The two \OH profiles of OH1 and OH2 are intergrated within a region with size of 0.0518906 $\times$ 0.08615 centered at RA: :20:57:24.375, Dec:17:07:39.131; The OH profile of OH2 was extracted from a region with size of 0.0639592 $\times$ 0.0837776 centered at RA:20:57:24.377, Dec:17:07:39.203.

  The uncertainties of integrated 3$\sigma$ flux density from each channel image is related to the area of the 3$\sigma$ pixels and the VLBI calibration error ($\sim$10\% of the measured flux): $\sqrt{(N \times \sigma)^{2}+(0.1 \times S_{peak})^2}$, where N represents the number of beams for the region of the 3$\sigma$ pixels and $S_{peak}$ is the peak flux density. The estimated uncertainties of the 3$\sigma$ profiles of OH1 and OH2 regions are about 4-5 mJy for the brightest channels and less than 1 mJy for the weak channels.  Based on these uncertainties, the OH profiles from the two epochs are consistent with each other and show no evident variabilities.

  We have also averaged the channels of OH 1667 MHz line emission from 10750 to 10950 \kms (see Fig. \ref{evnspectrum}) and imaged the OH1 and OH2 regions (see Fig. \ref{EVN2line}). We found that the peak position\textbf{s} of OH1 and OH2 show no evident changes between the two-epoch EVN line observations (see Table \ref{allline}).
  Because the OH 1665 MHz line emission is resolved with high-resolution observations, we have averaged the channels for this line emission (V $\sim$ 11080-11320 \kms, see Fig. \ref{evnspectrum}) and combined the calibrated data from the two EVN projects (using task "DBCON" in AIPS). As a result, the OH 1665 MHz line emission is detected at 6 $\sigma$ level with a peak of about 0.42 mJy/beam distributed at the OH1 region (see Fig. \ref{EVN2line}).

\subsection{The high-resolution CO (3-2) and HCO+ (4-3) line emission of this galaxy}
The HST I-band image (see Fig. \ref{radio-hst}) has shown that the OH emission region coincides with D1, surrounded by nuclei A, B, together with some other bright regions. We have selected 12 bright spots from the I-band image, named A1-A3, B1-B3, D1-D3, C1-C3, and we then extracted the CO (3-2) and HCO+ (4-3) spectra at these spots from the primary beam (PB) corrected image files %($\rm IIZw96\_CO\_image.pbcor.fits$, $\rm IIZw96\_HCOp\_image.pbcor.fits$ ) which is released online product
available online for the ALMA project 2012.1.01022.S (see Table \ref{vlbilist}).  %with beam size similar to the HST I-band image ($\sim$0.13" at 833.5nm) }.

We found that D1 is associated with the brightest CO (3-2) line emission ($\sim$ 153 mJy/beam) among all selected regions (see Fig. \ref{COabcd} and \ref{COall}), and at least ten times brighter than other spots in the peak CO(3-2) emission (e.g. A0 $\sim$ 6.35 mJy/beam, B0 $\sim$ 13.80 mJy/beam, see Table \ref{allline} and \ref{COgauss}). We further extracted the CO spectra from six regions around D1 (see Table \ref{COgauss} and Fig. \ref{COall}), and noted that all the selected areas showed CO spectra with peak velocities higher than the D1 component. The CO (3-2) emission in regions 1, 2 and 3 is much brighter than regions 4-6, which means that the southeast region might have higher CO densities at high velocities ($\sim$ from 10935.5 \kms to 11023.9 \kms)  as compared to the north-west region. We further present the velocity structure of the CO emission around the D1 component (see Fig. \ref{co-hco1115}); and confirmed that some regions show high-velocity clouds, while other areas show clear double peak emission lines (DP regions, DP1-DP6). The CO spectra in these regions are presented in Fig. \ref{COall}.

 We find that the HCO+ line can only be detected in the D1 region (see Fig. \ref{HCOabcd}), and the velocity map shows that some pixels at the edge exhibit slightly higher peak velocities (see Table \ref{allline} and Fig. \ref{co-hco1115}). %The contour plot of the HCO+ emission is present in Fig. \ref{cohcooh33}.
We have found that the CO and HCO+ line profiles extracted at the central region around D1 (with a size of $\sim$ 400 mas) all show broad FWHM (> 100 \kms) features, which are much wider than in other regions around this merging system.
%Because HCO+ is also a good 'dense gas' tracer \citep{2020MNRAS.499....1F},

\subsection{The \HI emission from VLA project AG0613 (see Table \ref{vlbilist})}
The \HI channel image of IIZw 096 (at velocity about 10849.3 \kms) is presented in Fig. \ref{vlaHI}. We see that there are two bright \HI emission regions: one is IIZw 096, the other is a new galaxy centered at RA: 20 57 39.307 and Dec: +17 01 53.762. The optical counterpart is likely to be SDSS J205738.81+170151.0, based on their celestial coordinates, as there is no optical redshift of this galaxy. The \HI profiles of the two galaxies and spots in IIZw 096 are presented in Fig. \ref{HItotal} and \ref{HI-region1-7}. The majority of the \HI emission (3 $\sigma$) for IIZw 096 is distributed in a region with a size of about 70"$\times$70" (see Fig. \ref{vlaHI}) and the integrated \HI spectrum in this region (see Fig. \ref{HItotal}) agrees well with the \HI spectrum from Arecibo observations by \cite{2015MNRAS.447.1531C}. The \HI spectra from VLA and Arecibo observations show that the galaxy contains both \HI emission and absorption. The \HI gas in absorption is detected only corresponding to the D1 component (see Fig. \ref{HItotal}), and this result is consistent with the GMRT observations made by \cite{2019MNRAS.489.1099D}.  The \HI absorption feature is at the highest end of the \HI spectrum velocity ranges. We have found that the \HI spectra from spots around IIZw 096 shows similar characteristics (see Table \ref{Regions-HI} and Fig. \ref{HI-region1-7}) and exhibit no evident orbiting velocity structures caused by the circular motions of the gas.
%The \HI absorption gas is distributed in a region about the size of the continuum emission 2 arcsecond around D1 \citep{2019MNRAS.489.1099D}, and it is consistent with the distance of the regions to D1 as shown in Fig. \ref{radio-hst}.
%which indicate that the \HI gas in absorption might be distributed around D1 and also moving inward to the center of D1.

\subsection{The high-resolution radio continuum emission}
  We have collected one epoch of VLBA data (see Table \ref{vladata}) for obtaining the high-resolution images of the continuum emission. The results show that no significant continuum emission detection occurs at a noise level of about 14.8 to 23.2 $\mu$Jy/beam (see Fig \ref{vlbacontinuum}).

  The VLA projects of this source and the measured radio flux densities of D and A are listed in Table \ref{vladata}, with the radio maps overlaid on the HST image of this source presented in Fig \ref{L-Ka1}. The data reduction and flux densities measurement methods are presented in Section 2.2.  Figs. \ref{ConstantABD} and \ref{ConstantDA-peak} show the multi-band radio spectra of D and A from the total and peak flux densities (see Table \ref{vladata}). We have used two models, one employs a single power-law and the other uses a mixed equation of single power-law and free-free emission ($S_{th} $$\times$ $\nu$$^{-0.1}$+$S_{nth}$ $\times$ $\nu$ $^{-\alpha}$), to fit the radio spectra. We see that the radio spectra of D cannot be well-fitted with a single power-law model; it is fitted with the mixed equation, which indicates that the source D might contain free-free emission and steep synchrotron emission.

\section{Discussion}{\label{sec:discussion}}
 As the detected OH line emission mainly originated from the D1 component of this merging system, we further investigated the high-resolution CO, HCO+ and \HI line emission and radio continuum emission of D1 and some other regions of this source (see Section 3). A combination of these properties might be helpful in analyzing the possible scheme associated with the OH emission in this source.
\subsection{The total mass of D1 estimated from CO and HI observations}

The mass contained in a region can be an indicator of whether an AGN might be present. The D1 region has been investigated in detail and we list here both our mass estimates of D1 and those published in the literature.

1. Based on H-band optical image and solar metallicity, \cite{2010AJ....140...63I} estimated that the mass of D1 is approximately 1–4 $\times$ $\rm 10^{9}$$\rm M_{\odot}$.

2. Based on the OH line velocities and emission region from MERLIN observations, \cite{2011MNRAS.416.1267M} obtained a lower limit for the enclosed mass of about 3 $\rm \times10^{9}M_{\odot}$. %\textbf{ There are all the estimates consistent with an AGN or could the masses also be consistent with starclusters}

3.
%\cite{2000astro.ph..8114G} have detected huge molecular gas concentrations around D1 region.
%\cite{1991ApJ...370..172P}
The $\rm H_{2}$ masses can be derived from CO (1-0) fluxes detected by the interferometer with the following equation \citep{2017ApJ...836..130R,1991ApJ...370..172P}:
 %$\rm M(H_{2})$=1.2$\times10^{4}\rm d^{2}_{Mpc}\int S_{Jy}$\, d$\upsilon$,
% $\rm M_{\rm H_{2}}$=1.05$\times$$10^{4}$$\frac{$\rm X_{CO}$}{2 $\times$ %$10^{20}$ $cm^{-2}$} $ \rm d^{2}_{Mpc}\int S_{Jy}$\, d$\upsilon$/(1+z),\hbox{$\psqcm(\K\kmps)^{-1}$}
 \begin{equation}
\rm M_{H_{2}} = 1.05\times 10^{4} \frac{\rm X_{CO}}{2\times 10^{20} \frac{\textrm{cm}^{-2}}{\textrm{K km s}^{-1}}}\frac{\rm D_{Mpc}^{2}}{1+z}\int S_{Jy}\, d\upsilon
\end{equation}
 where $\rm X_{CO}$ is the CO-to-$\rm H_{2}$ conversion factor in units of $\frac{\textrm{cm}^{-2}}{\textrm{K km s}^{-1}}$, z is the redshift, $\rm S_{Jy}$ is the CO(1-0) flux in Jy, $\rm d_{Mpc}$ is the luminosity distance of this object in Mpc. The CO(3–2)/CO(1–0) line ratio is about 0.51 for this galaxy \citep{2010MNRAS.406.1364L}, and likely, the LIRGs, Nearby Star-forming Galaxies, and Active Galactic Nuclei, all show similar line ratios \citep[see][]{2010MNRAS.406.1364L,2020ApJ...889..103L}. We adopted the $\rm X_{CO}$ = 0.4 $\times$ $10^{20}$ $\rm cm^{-2}$ $(K km^{-1})^{-1}$ for starburst galaxies and ULIRGs used in \cite{2017ApJ...836..130R,1998ApJ...507..615D}, which is about five times lower than the Galactic value.

 The $\rm d_{Mpc}$ is about 148 Mpc for this source, the $\int S_{Jy}$\, d$\upsilon$ is about 29175.5 mJy \kms /beam (see Table \ref{allline}), and the derived $\rm M(H_{2})$ is about 2.5$\times$ $10^{9}$ $\rm M_{\odot}$ in one beam with size of 0.2" $\times$ 0.16"  ($\sim$ 134 pc $\times$ 107 pc). Since the CO line emission of source D1 is much brighter than B0 and A0 (see Table \ref{allline}), the central mass in one beam for B0 and A0 are about 1.6$\times$ $10^{8}$ and 2.0$\times$ $10^{7}$ $\rm M_{\odot}$, respectively, which means that D1 possibly contains a much more massive central mass than the other two apparent nuclei.
%\textbf{1.5$\times$ $10^{10}$} $M_{\odot}$  , which shows the apparent brightest peak and %integrated line emission (see section 3.2). The $d_{Mpc}$ is about 148 Mpc for this source,

4. In the optically thin case, the \HI mass ($\rm M_{HI}$) can be derived from the integrated line flux with the equation:
    $\rm M_{HI}=2.36\times10^{5}\rm d^{2}_{Mpc}\int S_{Jy}\, d\upsilon$, where $\rm S_{Jy}$ is the line profile in Jy, integrated over the Doppler velocity V in \kms.  The $\rm \int S_{Jy}$\, d$\upsilon$ for the \HI spectrum of D1 is about 300.58 mJy \kms /beam (see Table \ref{allline}), whereby the \HI mass are about 1.6$ \times$ $10^{9}$ $\rm M_{\odot}$ in one beam with a size of 19.7" $\times$ 17.9" ($\sim$ 13.2 kpc $\times$ 12.0 kpc). Similarly, the estimated \HI mas from the total \HI spectrum (see Table \ref{allline}) is about 1.2$ \times$ $10^{10}$ $\rm M_{\odot}$ for the total \HI mass in IIZw 096.

All the above four estimated masses through different methods indicate that there is a massive mass of about $10^{9}$ $\rm M_{\odot}$ concentrated in the central region of D1. The direct proof of the existence of a super-massive black-hole requires very high angular resolution to probe close to the Schwarzschild radius, which has been difficult to establish \citep[see][]{2005ARA&A..43..625L}.  Alternatively, the large concentrations of molecular gas might result in high gas surface densities, which may signify that a luminosity source other than star formation, e.g., an AGN \citep{1999AJ....117.2632B}.

Based on the derived $\rm M(H_{2})$ and the physical size of one beam from CO (3-2) line observation (see the third estimation), the surface density is estimated to be about 2 $\times$ $10^{5}$ $\rm M_{\odot}$ $\rm pc^{-2}$, which is consistent with the value of Arp 220 as determined by \cite{2015ApJ...799...10B}. The high gas surface density value resembles the maximum stellar surface density of $\sim$ $10^{5}$ $\rm M_{\odot}$ $\rm pc^{-2}$ \citep{2010MNRAS.401L..19H}. It is believed that the most massive nuclear star clusters can reach mass surface densities of $\sim$ $10^{6}$ $\rm M_{\odot}$ $\rm pc^{-2}$ or more \citep[see][]{2020A&ARv..28....4N}. It is to be noted that the high surface gas density found in this source appears still below the highest mass surface densities in nuclear star clusters.  Meanwhile, the excessive-high gas surface density is also consistent with the view that there is a possibility it also hosts an obscured AGN, which might be responsible for the formation of an $\rm H_{2}O$ megamaser in this galaxy \citep[see][]{2016ApJ...816...55W,2011MNRAS.416.1267M}.

  %Based on the infrared luminosity, \cite{2010AJ....140...63I} estimated a large luminosity density of ∼4.5 \times$ $10^{12}$  $\rm L_{\odot}$ $\rm kpc^{−2}$ and this value is
 % .some probability of being a heavily obscured AGN  there is a possibility it also hosts an obscured AGN \cite{2011MNRAS.416.1267M}.
  %}

%and take the size of ource D estimated from the NICMOS image %(∼220 pc radius) and got a a luminosity density becomes %increasingly dominant as the gas surface density increases.  %\cite{1999AJ....117.2632B} show that the high gas surface %densities contain either more efficient or high-mass biased %star %formation, or else itneed the mass to radius ratio to be %equal to 6.7 $\times$ $10^{27}$ g/cm %\citep[see][]{2005ARA&A..43..625L}, which
\subsection{Comparison of OH and other line emissions}
\subsubsection{The merging and inflow of gas to the central region}
The merging or interaction between two or more galaxies can reduce the angular momentum of the circumnuclear material and enhance the inflow of material from galactic scales into the close environments of AGN \citep{2021MNRAS.506.5935R,2005Natur.433..604D}.
We found that the \HI profiles extracted from locations away from the center of this merging system (regions 1,2,4,5,7, and A0, see Fig. \ref{HI-region1-7} and Table \ref{Regions-HI})  show similar \HI profiles.  They may contain two components, with central velocities about 10760 \kms and 10840 \kms, respectively.  The optical redshift of this source shows a system velocity of about 10770 \kms \citep[e.g.][]{2019MNRAS.489.1099D,1995ApJS...98..129K}, which is consistent with the low-velocity \HI component.
The similar components for \HI emission profiles on a large scale at various regions indicate that this galaxy is in a stage of ongoing merging.  It suggests that the \HI gas clouds in this system might be distributed in a common face-on envelope and that the orbiting velocity cannot be measured through this observation, which is also observational evidence for intermediate merger \citep[e.g.][]{2021MNRAS.506.5935R,2011AJ....141..100H}.
The high-velocity line component might also be related to the gas clouds moving towards the center caused by the merging process.

%Then, the similar high velocity component exist in large regions might be the observational evidence of strong inflows towards the central region of this system rather than the orbiting velocity structures. , obscuring and feeding the supermassive black hole (SMBH), thus providing an effective mechanism to trigger accretion onto SMBHs
%It is likely that the \HI gas clouds at large scale are distributed in a face-on disk-structure that the orbiting velocity can not be measured through this observation and

%show that the cz about 10770 \kms and 10630 \kms for the NW and SE respectively. centered at 20 55 05.3 +16 56 03.Because the \HI emission is resolved with a resolution about 2" of GMRT observations,  then the \HI clouds for emission might mainly located at large scale, far from mass center of this system.

The regions HI6, D1, and B0 show two-components with slightly higher velocities (about 10780 \kms and 10880 \kms) than other regions (see Fig. \ref{HItotal} and \ref{HI-region1-7}, and Table \ref{Regions-HI}). Since these regions are close to the central region of this merging system, which contains massive nuclei (B0) and a possible supermassive region (D1) (see section 4.1), the velocity offset with other regions might be caused by the velocities directed toward the center of mass. The \HI emission is completely resolved with GMRT observations with a resolution of about 2" \citep{2019MNRAS.489.1099D}. The \HI absorption profile shows higher velocities than the \HI emission lines (see Fig. \ref{HItotal}). The detected \HI gas clouds in absorption were detected towards D1, which indicates that they may be closer to the central D1 than the large scale \HI gas in emission.
The velocity of the \HI absorption profiles is consistent with the high-velocity clouds seen from the velocity structure of CO emission (see Section 3.2). The high-velocity \HI and CO clouds around D1 might be related to the inflows towards the central region of this source.

%The \HI spectrum of this source show that the \HI absorption is at high end velocities ranges, which is consistent with the result from GMRT observations by \cite{2019MNRAS.489.1099D}, which show that the \HI abosrtion is towards D component, and there is an redshift velocity shift with systemic redshift about 144 km/s. The redshiftted \HI gas in absorption is likely detected toward D,
%it is consistent with the the results from CO emission regions around this component which all show redshiftted feature ( see section 3.2), which mean that the gas clouds around this component is likely moving inward to the center of this component, e.g. accreacting by the central supermassive black hole.

\subsubsection{The formation of OH line emission}

Two epoch EVN observations of the OH 1667 MHz line have confirmed the results from \cite{2010PhDT.......280C} that the high-resolution OH megamaser emission is detected from two regions (OH1 and OH2 as shown in Fig. \ref{radio-hst}). The OH1 emission is about two times brighter than OH2 (see Table \ref{allline}). We further found no evident variation of the OH properties from the two epoch observations (see section 3.1).
\cite{2008A&A...477..747B} show that the high-density tracers represent the molecular medium in the regions where star formation is taking place, which indicates a strong relationship to the far-infrared luminosity. In particular, the CO(3–2) transition is commonly used to trace the warmer, denser components of the ISM associated with star formation \citep{2010MNRAS.406.1364L}. The HCO+ is also believed to be a good 'dense gas' tracer \citep{2020MNRAS.499....1F}.

We have presented the results from the high-resolution observational data of OH 1667 MHz, CO(3-2), HCO+, and HI emission.
%The central velocity of OH1 and OH2 is about 10886 and 10795 \kms, respectively (see Table \ref{allline}).
We have found that the two OH emission regions (OH1 and OH2) reside in a dense gas environment (see Fig. \ref{co-hco1115}).
The central velocity of OH2 ($\sim$ 10804 \kms) is consistent with the peak velocity of the total \HI emission spectra (see Table \ref{allline} and Fig. \ref{HItotal}).
The central velocity of OH1 ($\sim$ 10886 \kms) is consistent with the dense gas tracer CO (3-2) and HCO+ of the D1 component. The OHM emerges when mergers experience a tidally driven density enhancement \citep{2007ApJ...669L...9D}.  Since the environment from CO (3-2) emission around D1 shows three regions with different velocity structures (see section 3.2), one possible scheme is that this region is in a merging stage. The OH1 and OH2 also emerge from the merging process and possibly originate from two or more systems. Our results show that the OH indeed is associated with the densest molecular gas regions found in the IIZw 096 system, agreeing with a scenario where star-formation is crucial for the formation of OHMs \citep{2005ARA&A..43..625L}.
%The OH masing regions are largely believed to be star forming \citep{2021A&ARv..29....2P}. }
%The the OH2 might originate from the OH clouds at system velocities, the OH1 is associated with other dense gas caused by merging process. In addition to the thick circumuclear structure, \cite{2007IAUS..242..446P} showed another possibleity is that the compact OH megamaser is the result of amplification in a clumpy, turbulent medium.

%These OH maser emission from OH1 and OH2 are caused by the on-going merging process, which are associated with the dense gas clouds which are proposed to be the trigure of OH megamaser emission \citep{2007ApJ...669L...9D}.

\subsubsection{Comparison with a general picture of OH megamaser emission}

The Ka-band VLA-A observations of this source (see Fig. \ref{radio-hst}) present the highest resolution radio continuum emission. We note that the CO(3-2), HCO+, and Ka-band VLA-A radio continuum emission are roughly coincident with each other, while the two OH emission regions (OH1 and OH2) arise from a location about 50-76 mas to the center of the 33 GHz emission (see Fig. \ref{cohcooh33}).

\cite{2009ASPC..407...73B} showed that an FIR radiation field from the dust could pump the OH molecules in an environment with n($\rm H_{2}$) in the range $10^{3}$ to 2 $\times$ $10^{4}$ $\rm cm^{-3}$. The maximum density of an OH maser emitting gas is of the order n($\rm H_{2}$)=$10^{5}$ $cm^{-3}$, while higher densities will thermalize the
energy levels and quench the maser emission \citep{2005A&A...443..383P}. We have calculated the n($\rm H_{2}$) using the following assumptions,  n($\rm H_{2}$) = $\rm \frac{M(H_{2})}{m(H_{2}) \times V}$ is about 4$\times$ $10^{4}$$\rm cm^{-3}$, where $\rm M(H_{2})$ is about 2.5$\times$ $10^{9}$ $\rm M_{\odot}$ and V is the volume of the central region in one beam with size of about 130 pc in diameter (see Section 4.1).  Since the accuracies of these contour maps are all less than 3.3 mas (estimated from $\rm \frac{beam}{S/N}$, where beam stands for the beam FWHM, S/N stands for the signal/noise ratio), the offset might be related with the high n($\rm H_{2}$) in the central region.

%the M($H_{2}$) is about 1.5$\times$ $10^{9}$ $M_{\odot}$ distributed in one beam regions , then the rough estimate of the , and found that both compact and apparently diffuse emission might be explained by similar clumpy structures. Speculative explanations for the lack of compact emission have been proposed (Lonsdale et al. 2003), but a simpler explanation is based on the geometry of a rotating clumpy maser disk.

%\subsection{Relations with existed OH megamaser models}
  Generally, high-resolution observations of the OH megamaser emission will find a velocity gradient across the region with two or more OH emission components, e.g., Arp 220~\citep{2009AJ....138.1529U}, III Zw 35~\citep{1997ApJ...485L..79T}, IRAS 17208-0014 \citep{2006ApJ...653.1172M}, and IRAS 12032+1707~\citep{2005ApJ...618..705P}, which is a sign of circular rotating disc or torus. Based on these high-resolution imaging of the 1667 MHz emission,  \cite{2007IAUS..242..446P}  presented a general picture where most of the maser emission arises in thick circumnuclear structures. \cite{2005A&A...443..383P} and \cite{2008ApJ...677..985L} showed that a clumpy maser model could provide a phenomenological explanation for both compact and diffuse OH emissions. Each maser cloud produces a low-gain, unsaturated emission; while compact emission would be observed when the line of sight intersects many maser clouds. This particular model has explained several OHM sources, e.g., III Zw 35, IRAS 17208-0014, Mrk 231, NGC 6240 \citep{2008ApJ...677..985L}.

MERLIN observations of IIZw 096 found weak evidence for a velocity gradient along the right ascension direction \citep{2011MNRAS.416.1267M}. However, the EVN observations showed that such a velocity gradient was a sign of double structure, and in this case no velocity gradient could be detected \citep[see][]{2010PhDT.......280C}. We also found that IIZw 096 also showed no clear presence of a velocity gradient from CO and HCO+ velocity distributions around D1 (see Fig. \ref{co-hco1115}). Since the velocity fields of OHMs might only become ordered during the final stage of merging \citep{2020A&A...638A..78P},  this specific source might still be dominated by a phase of intense merging, as seen from the optical images, rather than an ordered circumnuclear disk or torus. Such a scheme agrees very well with the view provided by \cite{1997AJ....113.1569G} that this source contains very young starbursts seen prior to the final major-merger stage.

%Based on the \cite{1997AJ....113.1569G}  suggests that starbursts in this galaxy are younger than those in most other infrared-luminous galaxies from the near-infrared spectroscopic diagnostics. these three galaxies (IIzw 096, Arp 299, VV 114 ) might represent a stage through which many, perhaps most, luminous infrared galax- ies will pass.and more likely to be luminous “intermediate” stage starbursts from the morphological and spectral characteristics.which contain two nuclei in common envelope
Generally, the merger classification scheme is based on the morphology obtained from the high-resolution optical or infrared images. IIZw 096 was classified to be at the intermediate stage of the merging process based on the reported morphological characteristics \citep[e.g.,][] {1997AJ....113.1569G,2011AJ....141..100H,2021MNRAS.506.5935R}. \cite{2016ApJ...816...55W} have shown that galaxies with coexisting OH and water megamasers might be a distinct population at a brief phase along the merger sequence due to the independent natures of the two types of maser emission. Although extensive effort has been made to detect OH+water megamasers, IIZw 096 and Arp 299 are the only two galaxies that are confirmed with such a dual-megamaser \citep[see][and references therein]{2016ApJ...816...55W}. The similarities between IIZw 096 and Arp 299 have shown that they are both luminous intermediate stage starburst \citep[see][]{1997AJ....113.1569G,2010AJ....140...63I,2016ApJ...816...55W}. The evidence of inflow found in the two galaxies \citep[see][and Section 4.2.1]{2017A&A...597A.105F} might also be a similarity related to their analogous merging stage. Since strong HCN and OH emissions are both detected in the source IC 694 of Arp 299 \citep{1999A&A...346..663C,2002IAUS..206..430K}, it further confirms that the OH megamaser emission is associated with the dense gas environment in the two galaxies. One significant difference between the two galaxies is in their velocity structure: a regular velocity gradient is seen around IC 694 from \HI, CO, and OH line emission \citep{2000evn..conf..127P,2001IAUS..205..198P,1999A&A...346..663C}, while we found no velocity gradient in IIZw 096. An explanation given by \cite{1991ApJ...366L...1S} is that Arp 299 is at a more advanced state of merging, and IC 694 is well on its way to becoming the ultimate core of the merger. Although the two galaxies might be both at the intermediate stage of merging, the absence of a velocity gradient means that IIZw 096 might be experiencing a slightly earlier stage of merger as compared to Arp 299.

\subsection{The radio continuum}
 The OHM galaxies could either represent a transition stage between a starburst and the emergence of an AGN through the merging process \citep{2020A&A...638A..78P} or harbor a recently triggered AGN \citep{2020MNRAS.498.2632H}. IIZw 096 is the second object to co-host both water and OH megamasers, and these dual-megamasers might be only emerging during a brief phase of the galaxy evolution \citep[e.g., from starburst nucleus to an AGN, see][]{2016ApJ...816...55W}. \cite{2011MNRAS.416.1267M} first proposed a possibility that IIZw 096 might host an obscured AGN. However, multi-band observations reported in the literature show no clear
 evidence for an obscured AGN: the X-ray spectrum can be well-fit using the star-formation mode \citep{2021MNRAS.506.5935R,2011A&A...529A.106I}; and the Spitzer mid-infrared spectra indicate no high-ionization lines from a buried AGN \citep{2010AJ....140...63I}.

%Because non-AGN ULIRGs at optical wavelengths show similar radio properties to those of the entire ULIRG sample, \cite{2021MNRAS.504.2675H} imply that AGNS might equally contribute to the radio emission of every ULIRG.  there is a possibility  as firstly proposed by \cite{2011MNRAS.416.1267M},

\cite{2006A&A...449..559B} have classified this source as a starburst galaxy, based on the radio brightness temperature ($T_{b}$) from VLA observations, the radio spectral index, and the ratio of FIR and radio flux.
Since the high-resolution radio structure and $T_{b}$ are extremely significant ways to distinguish AGN from SB galaxies \citep{1991ApJ...378...65C}, %There is giving the $T_{b}$ formula $T_{b}$$=$1.66$\times10^{12}$$(S_{\nu})$$\nu^{-2}$$\theta_{d}^{-2}$(1+z) \citep{1991Natur.354..130Y}.
 we have investigated the 33 GHz (Ka-band)  VLA-A observation of this source (project 14A-471, see Table \ref{vlbilist}) and model-fitted the visibility data. Assuming the flux density of the fitted component can also be obtained at 1.4 GHz with the same component size, the estimated temperature $T_{b}$ of the radio continuum emission is ~1.21$\times$$10^{5}$ K. The estimated upper limit $T_{b}$ from the peak flux densities of VLBA images (as shown in Fig. \ref{vlbacontinuum}) is about $10^{6}$ K. Thereby, the detected radio continuum emission from comp D of IIZw 096 show\textbf{s} $T_{b}$ to be within the range $10^{5}$ - $10^{6}$ K, which is also consistent with the starburst origin radio emission of other OHM galaxies reported in the literature \citep[e.g., IRAS 12032+1707, IRAS 02524+2046 in][]{2005A&A...443..383P,2020A&A...638A..78P}.

\cite{2008A&A...477...95C} have shown that very few LIRGs/ULIRGs have a straight power-law slope, where the overall radio SED begins to flatten at higher frequencies as the contribution of thermal emission increases. We have fitted the multi-band radio continuum of the comp D and A as presented in Section 3.4. The radio SED of A can be well-fitted with the power-law equation, while comp D might contain contributions from the free-free emission, which might indicate the existence of \HII regions related to the massive stars \citep{2019ApJ...881...70L}.  \cite{2015A&A...574A...4V} have shown that if the starburst is situated in an HII region, the value of $\alpha$ is hardly steeper than  1.1 from the $\alpha$-map when adopting a 2 $\sigma$ uncertainty, which is an effective way to classify the LIRGs as radio-AGN, radio-SB, and AGN/SB (a mixture).

%  found that spatial variations of  $\alpha$ from the radio-spectral-index maps is an effective way to classify the LIRGs as radio-AGN, radio-SB, and AGN/SB (a mixture).

%The radio spectra is one critical parameter of classification of the radio activity in the nuclei, other two parameters are radio brightness temperature, and the ratio of FIR and radio fluxes \citep{2006A&A...449..559B}. Generally, AGN-dominated sources have intrinsically flat ($\alpha$$\leq$0.5) radio spectra, moderately radio spectra by themselves cannot be used to rule out starbursts among the luminous FIR galaxies, however, steeper spectra do indicate significant starburst contributions. The median spectral index of normal galaxies and extended starbursts is $\alpha$$\approx$0.75 \citep{2006A&A...449..559B}.

 %\subsection{The spectral index map from VLA data}
 % \cite{2015A&A...574A...4V} have classified the LIRGs to contain radio-AGN when $i$ $\alpha$ has negative values (inverted) due to synchrotron self-absorption from a jet pointing towards the observer; $ii$) $\alpha$ is flat in the centre, displaying a steep tail suggesting synchrotron ageing, and the values of $\alpha$ get larger than 1.1; $iii$) the $\alpha$-map has steep $\alpha$ values that exceed 1.1; $iv$)

 Following the methods as descripted in \cite{2015A&A...574A...4V}, we have constructed the $\alpha$-maps (see Fig. \ref{359}). First, we restored the images (at 1.4 GHz, 3 GHz, and 9 GHz) to a beam size of 2" (comparable to the default beam of VLA-A observations at the L band) and constructed the $\alpha$-maps.  We found no regions with a steep spectral index higher than 1.1. Second, we restored the images at 3 and 9 GHz to a beam size of 1" (comparable to the default beam of VLA-B observations at 9 GHz, see Fig. \ref{vladata}) and re-constructed the $\alpha$-maps, whereby we found that there are some pixels with $\alpha$ steeper than 1.1.  Although this might be a sign of the possible existence of radio AGN in the central region, the number of pixels with $\alpha$ >1.1 is scarce based on the histograms as shown in Fig. \ref{vladata}. By combining properties of the radio continuum emission from brightness temperature and multi-band fitting, along with the properties of X-ray and infrared emission \citep{2021MNRAS.506.5935R,2011A&A...529A.106I,2010AJ....140...63I}, we conclude that the dominated radio continuum emission from D1 might be of starburst origin, contributing to both synchrotron and free-free emission.

\section{Summary}
IIZw 096 is likely to be a rare nearby example, potentially for the study of merging environments and OH megamaser emissions. We have analyzed two-epoch EVN archival data of the OH 1667 MHz line emission of IIZw 096 and confirmed that this source's OH 1667 MHz line emission is mainly from two regions. We found no significant variations of the OH 1667 MHz line emission from the two areas, including the integrated flux densities and peak positions. The OH 1665 MHz line emission is detected at the 6$\sigma$ level, with a peak of about 0.42 mJy/beam from the OH1 region.
 The OH emission regions reside in comp D1, which shows the brightest CO and HCO+ emission. The molecular mass in the central part ($\sim$~130 pc) is about 2.5$\times$ $10^{9}$, which is consistent with the view that there is a high mass concentrated in the central region \citep{2011MNRAS.416.1267M,2010AJ....140...63I}.
%might be over $10^{10}$ $M_{\odot}$as defined by \cite{2021MNRAS.506.5935R}

 The \HI emission from the VLA data shows that the \HI gas at large scales may be distributed in a common face-on envelope, which is consistent with a stage of intermediate merger. The CO velocity structures show
 three-velocity structures around D1: 1) a broad line profile region , which is the central region around D1 where CO emission shows broad line profiles; 2) a double peak region, which has several small areas surrounding the central region where the CO line profile showed double peaks; 3) a high-velocity clouds region, further out of the double-peak regions, where the CO spectrum reveals high velocities around 11000 \kms.  One possible explanation is that this source is in a stage of ongoing merging of two or more systems. The velocity structure around D1 shows no evidence of  circular motions, making it different from most other OHMs reported in the literature, which might be caused by an effect resulting from the merger stage. The CO, HCO+ line emission, and the K-band VLA-A continuum emission are roughly aligned with the brightest center; while the two OH emission regions show an offset of about 50-75 mas to the central region and tend to the direction of double peak region. Thereby, the two OH emission regions might also be related to the merging process and may originate from more than one system.

We found that there is no significant continuum emission from the VLBA archival data. The multi-band radio continuum emission shows that the radio SED of comp A can be well-fitted with the power-law equation, while comp D might contain contributions from the free-free emission.  The $\alpha$-map shows regions steeper than 1.1,  which might be a sign of the possible existence of radio AGN in the central part as reported by \cite{2015A&A...574A...4V}. However, the pixels steeper than 1.1 are very rare; and it is likely that the dominated radio continuum emission has a starburst origin mixed with synchrotron and free-free emission.

  %II Zw 096 showed a complex characteristics, it did not show significant radio continuum at higher resolution. In II Zw 096, the compact activity that is not within the nuclear regions of the merger.  two regions which are likely show no velocity gradient related to the circular motions. there are also regions show double peaks for the detected spectrum, which is located

  %we think source D can be taken to be a galactic nucleus.

\begin{acknowledgements}
  We thank the referee for the constructive comments and suggestions, which helped improve this paper.
  The study was funded by RFBR and NSFC, project number 21-52-53035 ``The Radio Properties and Structure of OH Megamaser Galaxies''.
  This work is supported by the grants of NSFC (Grant No. 11763002, U1931203).
  The European VLBI Network is a joint facility of European,
  Chinese, and other radio astronomy institutes funded by their national research councils.
  The National Radio Astronomy Observatory is operated by Associated Universities,
  Inc., under cooperative agreement with the National Science Foundation. This paper makes use of the following ALMA data: ADS/JAO.ALMA\#2012.1.01022.S. ALMA is a partnership of ESO (representing its member states), NSF (USA) and NINS (Japan), together with NRC (Canada), MOST and ASIAA (Taiwan), and KASI (Republic of Korea), in cooperation with the Republic of Chile. The Joint ALMA Observatory is operated by ESO, AUI/NRAO and NAOJ.

\end{acknowledgements}

\bibliography{MGJ04}

%-------------------------------------------------------------------
\appendix
\section{Online materials}
\twocolumn
\counterwithin{figure}{section}
%--------------------------------------------------------------------
  \setlength{\tabcolsep}{0.01in}
  \begin{table*}
       \caption{Parameters of the high\textbf{-}resolution radio continuum observations. }
     \label{vladata}
  \begin{center}
  \begin{tabular}{c c c c l c c c c c c c}     % 8 columns
  \hline\hline
    Observing Date & Frequencies & Array    & Phase      & Program     & beam &PA &rms &Component     & Integrated Flux       & Map peak               \\
                   &   (GHz)     & Configuration  &  Calibrator          &  & (") $\times$(")    & ($\circ$) & (mJy/beam) &    &   (mJy)               &(mJy/beam)              \\
      \hline
    1992Dec14      & 1.4        & VLA-A   &2029+121       & AB0660       & 2.86  $\times$1.54  &61& 0.43 &D          & 15 $\pm$ 2      & 6.9 $\pm$ 0.7    \\
                   &             &                &      &       &  & &&A          & 5.4 $\pm$ 0.5       & 3.6 $\pm$ 0.2    \\
    2014Mar30      & 3.0        & EVLA-A  &J2139+1423        & 14A-471     & 0.67 $\times$ 0.61&5& 0.040 &D          &7.0 $\pm$ 0.5       &4.9 $\pm$ 0.2    \\
                   &             &                &     &        &  && &A          & 3.2 $\pm$ 0.4      & 2.0 $\pm$ 0.2    \\
    2013Dec01      & 5.1        & EVLA-B  &...        & 13B-356       &1.50 $\times$ 1.42  &-41& 0.041& D          & 5.9 $\pm$ 0.5  & 4.1 $\pm$ 0.2    \\
                   &             &                &      &       & &&  &A          &2.6 $\pm$ 0.2     & 1.60 $\pm$ 0.09    \\
    2013Dec01      & 7.0        & EVLA-B  &...        & 13B-356     &1.10 $\times$ 1.07 &51&0.028 & D          & 4.8 $\pm$ 0.2       & 3.4 $\pm$ 0.1    \\
                   &             &                &      &       & && & A          & 1.9 $\pm$ 0.1       & 1.20 $\pm$ 0.06    \\
    2013Dec01      & 9.0        & EVLA-B   &...       & 13B-356     &0.96 $\times$ 0.79 &-58&0.023 & D          &4.3 $\pm$ 0.2       & 3.16 $\pm$ 0.08    \\
                   &             &                &      &       & && & A          & 1.77 $\pm$ 0.06       & 1.16 $\pm$ 0.03    \\
    2013Dec01      & 11.0       & EVLA-B  &...        & 13B-356     &0.83 $\times$ 0.65 &-60&0.027 & D          & 3.6 $\pm$ 0.2       & 2.63 $\pm$ 0.07    \\
                   &             &                &       &      & && & A          & 1.65 $\pm$ 0.07       & 1.03 $\pm$ 0.03    \\
    2013Nov15      & 13.0       & EVLA-B  &...        & 13B-356     &0.84 $\times$ 0.52&-2& 0.021 & D          & 3.5 $\pm$ 0.3       & 2.7 $\pm$ 0.1    \\
                   &             &                &      &       & &&  &A          & 1.33 $\pm$ 0.06       & 0.87 $\pm$ 0.03    \\
    2013Nov15      & 15.0       & EVLA-B   &...       & 13B-356     & 0.52 $\times$ 0.47 &28&0.022 &D          & 3.3 $\pm$ 0.1       & 2.53 $\pm$ 0.06    \\
                   &             &                &      &       & && & A          & 1.25 $\pm$ 0.08       & 0.84 $\pm$ 0.03    \\
    2013Nov15      & 17.0       & EVLA-B  &...        & 13B-356     &0.46 $\times$ 0.42 &36& 0.026& D          &3.3 $\pm$ 0.1       & 2.53 $\pm$ 0.06    \\
                   &             &                &     &        & && & A          & 1.02 $\pm$ 0.03       & 0.72 $\pm$ 0.02    \\
    2013Nov15      & 21.9       & EVLA-B   &...       & 13B-356     &0.39 $\times$ 0.33 &54&0.015 & D          & 2.94 $\pm$ 0.08       & 2.34 $\pm$ 0.04    \\
                   &             &                &       &      & && & A          & 0.91 $\pm$ 0.04       & 0.66 $\pm$ 0.02    \\
    2014Nov06      & 33.0       & EVLA-C  &...        & 14A-471     &0.96 $\times$ 0.72 &56&0.025 & D          & 2.9 $\pm$ 0.1       & 2.16 $\pm$ 0.06    \\
                   &             &                &      &       &&&   &A          & 0.80 $\pm$ 0.11       & 0.63 $\pm$ 0.06    \\
   2014Apr18      & 33.0       & EVLA-A &...         & 14A-741  &0.082 $\times$ 0.060 & 69& 0.015 &  -          & -      &-    \\
   2014May18/19      & 1.5       & VLBA &  J2052+1619$\rm ^{pr}$       & BS0233  &0.005 $\times$ 0.011 & 2 &0.015&  -          & -      &-    \\

   %2014Nov02      & 3.00       & VLA-C                & 14A-471     &   B          & 5.800 $\pm$1.4005  & 1.250$\pm$0.260   \\
%    2011Feb13      & 4.70       & VLA-C$\Rightarrow$B  & AL746       &   B          & 3.010 $\pm$0.8104  & 0.829$\pm$0.177   \\
 %         & 5.95       &   &      &   B          & 2.260 $\pm$0.6704  & 0.726$\pm$0.166   \\
  %        & 7.20       &   &      &   B          & 1.690 $\pm$0.5502  & 0.532$\pm$0.133   \\
   % 2014Oct31      & 12.49       & VLA-C$\Rightarrow$B  & 14A-471     &   B          & 0.966 $\pm$0.2702  & 0.369$\pm$0.077   \\
%          & 15.00       &   &      &   B          & 1.080 $\pm$0.3302  & 0.360$\pm$0.084   \\
 %         & 17.50       &   &      &   B          & 0.788 $\pm$0.2812  & 0.323$\pm$0.085   \\
 %   2014Nov06      & 33.00       & VLA-C                & 14A-471     &   B          & 0.762 $\pm$0.2811  & 0.212$\pm$0.062   \\
 %         & 35.00       &                 &      &   B          & 0.680 $\pm$0.2051  & 0.239$\pm$0.054   \\
 %   2001Jul22    & 1.37        & VLA-C          & AG0613*     &   -                 &  -     & -    \\
      \hline
       \end{tabular}\\
    %$\textbf{Note.}$ The Integrated Flux and the Map peak is from Common Astronomy Software Applications(CASA). We obtained the Integrated Flux and the Map peak and their errors by selecting a circle with 6.28434 $\times$ 6.28547 (width$\times$ length ), fitting it into component D at the central position of RA:20:57:24.375,   Dec:17:07:39.14. Similarly, we use the circle with 4.155 $\times$ 4.155(width$\times$ length ) to fit into component A at the central position of RA:20:57:24.078, Dec:17:07:39.14.
    \end{center}
     \vskip 0.1 true cm \noindent Column (4): The phase-reference calibrator for VLBA project and phase calibrator for VLA projects. The index 'pr' means the project is in phase-reference mode. Column (5): The program name. Column (6) and (7): The beam FWHM and position angle. Column (8): The 1 $\sigma$ noise level for the radio continuum image as present in Fig. \ref{L-Ka1}.  Column (9): The component name. Column (10)-(11): The integrated and peak flux densities measured from restored images with a beam size of 2" $\times$ 2", respectively.
    \end{table*}

  \begin{table*}
       \caption{The CO line spectrum of the components and regions in IIZw 096. \label{COgauss} }
     \label{components}
  \begin{center}
  \begin{tabular}{c c c c l c c c c}     % 8 columns
  \hline\hline
 Components(CO)  &  Opt. Coords (J2000)        & Gauss amplitude   &Gauss center        & Gauss FWHM         &Gaussian area       \\
             & (hh mm ss,+dd mm ss)        & mJy/beam          & \kms                 & \kms               & mJy/beam*\kms       \\
   D0:      & 20 57 24.340, 17 07 39.084  & 14 $\pm$ 1    & 10898 $\pm$ 1    & 53 $\pm$ 3    & 795 $\pm$ 46   \\
   D1:      & 20 57 24.372, 17 07 39.221  & 153 $\pm$ 1    & 10888 $\pm$ 1    & 179 $\pm$ 1    & 29176 $\pm$ 155  \\
   D2:      & 20 57 24.389, 17 07 40.030  & -2.1 $\pm$ 0.6    & 10965 $\pm$ 10    & 72 $\pm$ 22   & -162 $\pm$ 47   \\

   C0:      & 20 57 24.473, 17 07 39.826  & --                 & --                   & --                 & --                  \\
   C1:      & 20 57 24.519, 17 07 40.595  & --                 & --                   & --                 & --                  \\
   C2:      & 20 57 24.536, 17 07 40.840  & 0.9  $\pm$ 0.5     & 11291 $\pm$ 38   & 138 $\pm$ 90  & 129 $\pm$ 79    \\
   C3:      & 20 57 24.481, 17 07 40.881  & --                 & --                   & --                 & --                  \\

   B0:      & 20 57 23.604, 17 07 44.387  & 14 $\pm$ 2    & 10880 $\pm$ 8    & 124 $\pm$ 19   & 1821 $\pm$ 257  \\
   B1:      & 20 57 23.964, 17 07 41.631  & 2.4  $\pm$ 1.0    & 10929 $\pm$ 12    & 59 $\pm$ 28   & 147 $\pm$ 67   \\
   B2:      & 20 57 23.901, 17 07 42.001  & --                 & --                   & --                 &                     \\
   B3:      & 20 57 24.070, 17 07 40.494  & -2.3 $\pm$ 0.9    & 10631 $\pm$ 8    & 41 $\pm$ 20   & -99 $\pm$ 44   \\
   B4:      & 20 57 23.824, 17 07 42.411  & --                 & --                   & --                 & --                  \\
   B5:      & 20 57 23.757, 17 07 42.141  & -2.6 $\pm$ 1.3    & 11385 $\pm$ 20   & 80 $\pm$47   & -218 $\pm$ 119  \\
   B6:      & 20 57 23.796, 17 07 41.250  & 2.5  $\pm$ 0.9    & 10901 $\pm$ 25   & 149 $\pm$ 60   & 399 $\pm$ 105  \\

   A0:      & 20 57 24.069, 17 07 34.921  & 6.4  $\pm$ 1.2    & 10768 $\pm$ 3     & 34 $\pm$ 7    & 232 $\pm$ 46    \\
   A1:      & 20 57 24.167, 17 07 36.440  & --                 & --                   & --                 & --                  \\
   A2:      & 20 57 24.217, 17 07 36.654  & -0.8 $\pm$ 0.7    & 10929 $\pm$ 33   & 74 $\pm$ 76   & -60 $\pm$ 57    \\
   A3:      & 20 57 24.274, 17 07 38.119  & 8.6  $\pm$ 1.0    & 10904 $\pm$ 1     & 26 $\pm$ 3    & 236 $\pm$ 28    \\
 \hline
   DP 2     &          & 11 $\pm$ 2        & 10916 $\pm$ 5    & 66 $\pm$ 11      & 779 $\pm$ 175    \\   % 1
             &         & 3.7 $\pm$ 1.1         & 11018 $\pm$ 18   & 82 $\pm$ 35       & 342  $\pm$ 161      \\   % 2
     %        &         & 11.1$\pm$1.9        & 10916.2$\pm$9.9    & 68.4$\pm$19.8      & 1103.0$\pm$256.9     \\   % 1+2

   DP 3      &         & 10 $\pm$ 2         & 10920 $\pm$ 8    & 665 $\pm$ 16    & 687 $\pm$ 202    \\   % 1
             &         & 4.1 $\pm$ 1.5         & 11016 $\pm$ 21   & 81 $\pm$ 43     & 369 $\pm$ 215    \\   % 2
    %         &         & 10.1$\pm$2.0        & 10920.4$\pm$14.4   & 69.2$\pm$28.8      & 1055.4$\pm$321.8     \\   % 1+2

   DP 4      &         & 9.5 $\pm$ 2.0         & 10978 $\pm$ 7    & 50 $\pm$ 15     & 509 $\pm$ 173     \\   % 1
             &         & 4.9 $\pm$ 1.6         & 11064$\pm$ 11   & 27$\pm$21      & 142 $\pm$ 112    \\   % 2
    %         &         & 9.5$\pm$2.0         & 10977.7$\pm$7.4    & 50.0$\pm$14.9      & 650.4 $\pm$172.4     \\   % 1+2

   DP 5      &         & 6.8 $\pm$ 1.7         & 10931 $\pm$ 8    & 32 $\pm$ 17      & 231 $\pm$ 126    \\   % 1
             &         & 3.6 $\pm$ 1.5         & 10977 $\pm$ 28   & 99 $\pm$ 56      & 376 $\pm$ 275     \\   % 2
      %       &         & 8.8$\pm$1.9         & 10932.2$\pm$14.2   & 41.6$\pm$28.4      & 607.1 $\pm$216.9     \\   % 1+2

   DP 6      &         & 4.0 $\pm$ 1.2         & 10894 $\pm$ 17   & 77 $\pm$ 33      & 329 $\pm$ 164     \\   % 1
             &         & 3.0 $\pm$ 1.2         & 10984 $\pm$ 18   & 50 $\pm$ 36      & 161 $\pm$ 124     \\   % 2
     %        &         & 4.0$\pm$1.2         & 10893.9$\pm$26.9   & 79.6$\pm$53.7      & 489.8 $\pm$237.6     \\   % 1+2
     \hline
            & (hh mm ss,+dd mm ss)        & mJy          & \kms                 & \kms               & mJy*\kms       \\

   Reg. 1    & 20 57 24.386, 17 07 38.574  &  8.1 $\pm$ 0.4    & 11014 $\pm$ 2      & 93 $\pm$ 5       & 807 $\pm$ 42    \\
   Reg. 2    & 20 57 24.411, 17 07 38.734  &  10.8 $\pm$ 0.4    & 11024 $\pm$ 2     & 77 $\pm$ 4       & 881 $\pm$ 38    \\
   Reg. 3    & 20 57 24.423, 17 07 39.283  &  12.5$\pm$ 0.4    & 10962 $\pm$ 1      & 55 $\pm$ 2       & 734 $\pm$ 28    \\
   Reg. 4    & 20 57 24.306, 17 07 39.747  &  3.4 $\pm$ 0.5    & 10910 $\pm$ 3      & 41 $\pm$ 7       & 150 $\pm$ 24    \\
   Reg. 5    & 20 57 24.319, 17 07 40.093  &  -1.7 $\pm$ 0.2    & 10936 $\pm$ 13     & 178 $\pm$ 30      & -312 $\pm$ 50    \\
   Reg.6    & 20 57 24.360, 17 07 39.232  &  -1.5 $\pm$ 0.4    & 10942 $\pm$ 9      & 69 $\pm$ 22      & -106 $\pm$ 31    \\

 % DP 6-1    &         & 6.8$\pm$1.8         & 10902.1$\pm$10.3   & 44.2$\pm$20.5      & 321.18 $\pm$160.59     \\   % 1
 %            &         & 3.0$\pm$1.5         & 10986.0$\pm$22.7   & 42.1$\pm$45.3      & 135.20 $\pm$151.00     \\   % 2
 %            &         & 6.8$\pm$1.8         & 10902.1$\pm$10.4   & 44.4$\pm$20.7      & 456.38 $\pm$161.92     \\   % 1+2
  \hline
  \end{tabular}\\

    \end{center}
    \vskip 0.1 true cm \noindent Columns (1): the emission components; (2): RA and Dec of the components or spots; Columns (3)-(6) The Gaussian-fitted model parameters.
    \end{table*}
\begin{table*}
       \caption{Parameters of the HI emission line regions in IIZw 096}
     \label{Regions-HI}
  \begin{center}
  \begin{tabular}{c c c c l c c c c}     % 8 columns
  \hline\hline
 Regions(HI) & RA(J2000)  Dec(J2000)       &  Gauss amplitude     & Gauss center          & Gauss FWHM         & Gaussian area     \\
              &                            & mJy/beam             & \kms                  & \kms                 & mJy/beam*\kms           \\
  \hline
     HI1   %   & 20 57 25.251, 17 07 27.405  &  1.8$\pm$0.5         & 10823.8$\pm$19.8      & 152.8$\pm$39.7       & 268.99$\pm$108.26    \\
             & 20 57 25.251, 17 07 27.405                         &  1.4 $\pm$ 0.5         & 10765 $\pm$ 26      & 109 $\pm$ 52       & 160 $\pm$ 91      \\
             & --                          &  1.3 $\pm$ 0.5         & 10840 $\pm$ 23      &  79 $\pm$ 47       & 109 $\pm$ 73     \\

     HI2    %  & 20 57 24.589, 17 07 15.846  &  1.1$\pm$0.4         & 10829.3$\pm$23.1      & 170.0$\pm$46.2       & 190.01$\pm$82.60     \\
             &20 57 24.589, 17 07 15.846                         &  0.9 $\pm$ 0.4         & 10760 $\pm$ 27      & 96 $\pm$ 54       & 93 $\pm$ 61     \\
             & --                          &  1.0 $\pm$ 0.4         & 10846 $\pm$ 25      & 95 $\pm$ 51       & 97 $\pm$ 60     \\

     HI3    %  & 20 57 23.779, 17 07 31.670  &  1.5$\pm$0.5         & 10829.3$\pm$21.9      & 198.0 $\pm$43.8      & 314.44$\pm$113.42    \\
             &  20 57 23.779, 17 07 31.670                          &  0.7 $\pm$ 0.4         & 10769 $\pm$ 40      & 95  $\pm$ 79      &  71$\pm$68     \\
             & --                          &  1.5 $\pm$ 0.5         & 10856 $\pm$ 25      & 157 $\pm$ 50      & 243 $\pm$ 101    \\

     HI4    %  & 20 57 22.602, 17 07 53.824  &  1.8$\pm$0.4         & 10766.4$\pm$13.2      & 152.0 $\pm$26.4      & 260.03$\pm$79.44     \\
             & 20 57 22.602, 17 07 53.824                          &  1.6 $\pm$ 0.4         & 10759 $\pm$ 14      & 80  $\pm$ 27      & 137 $\pm$ 54     \\
             & --                          &  1.3 $\pm$ 0.5         & 10839 $\pm$ 17      & 87  $\pm$ 34      & 123 $\pm$ 56     \\

     HI5   %   & 20 57 24.110, 17 08 02.264  &  1.2$\pm$0.4         & 10745.3$\pm$21.4      & 187.6 $\pm$42.8      & 228.92$\pm$91.38     \\
             &  20 57 24.110, 17 08 02.264                          &  1.2 $\pm$ 0.4         & 10736 $\pm$ 22      & 103 $\pm$ 45      & 126 $\pm$ 65     \\
             & --                          &  1.0 $\pm$ 0.4         & 10834 $\pm$ 27      & 101 $\pm$ 54      & 103 $\pm$ 64     \\

     HI6   %   & 20 57 24.129, 17 07 45.385  &  2.1$\pm$0.9         & 10860.7$\pm$32.3      & 204.4 $\pm$64.6      & 443.85$\pm$323.98     \\
             & 20 57 24.129, 17 07 45.385                         &  2.1 $\pm$ 0.9         & 10778 $\pm$ 38      & 161 $\pm$ 76      & 357 $\pm$ 217     \\
             & --                          &  1.2 $\pm$ 0.9         & 10873 $\pm$ 41      & 66  $\pm$ 82      & 87 $\pm$ 117     \\

     HI7   %   & 20 57 25.601, 17 07 46.703  &  1.6$\pm$0.5         & 10828.6$\pm$35.1      & 24.0  $\pm$70.2      & 186.07$\pm$110.73     \\
             &  20 57 25.601, 17 07 46.703                          &  0.8 $\pm$ 0.4         & 10743 $\pm$ 41      & 124 $\pm$ 83      & 105 $\pm$ 85      \\
             & --                          &  1.4 $\pm$ 0.5         & 10831 $\pm$ 16      & 55  $\pm$ 32      & 81 $\pm$ 51      \\

   %  +D1   %  & 20 57 24.372, 17 07 39.221  &  1.8$\pm$0.5         & 10780.2$\pm$14.9      & 172.4 $\pm$29.8      & 318.58$\pm$98.38      \\
    %         &  20 57 24.372, 17 07 39.221                          &  1.8$\pm$0.5         & 10780.2$\pm$17.4      & 134.7 $\pm$34.8      & 246.92$\pm$90.42      \\
     %        & --                          &  1.0$\pm$0.4         & 10863.1$\pm$18.9      & 49.2  $\pm$37.9      & 53.66 $\pm$43.80      \\

     +B0   %  & 20 57 23.604, 17 07 44.387  &  2.3$\pm$0.4         & 10872.9$\pm$6.3       & 232.8 $\pm$12.5      & 545.85$\pm$92.31      \\
             & 20 57 23.604, 17 07 44.387                         &  1.9 $\pm$ 0.3         & 10787 $\pm$ 8        & 229 $\pm$ 17      & 466 $\pm$ 80     \\
             & --                          &  1.1 $\pm$ 0.2         & 10883 $\pm$ 8       & 69  $\pm$ 16      & 80 $\pm$ 23      \\

     +A0   %  & 20 57 24.069, 17 07 34.921  &  2.1$\pm$0.5         & 10774.0$\pm$12.8      & 147.2 $\pm$25.5      & 283.33$\pm$90.33      \\
             &  20 57 24.069, 17 07 34.921                        &  1.7 $\pm$ 0.5         & 10769 $\pm$ 13      & 54  $\pm$ 25      & 96 $\pm$ 49      \\
             & --                          &  1.7 $\pm$ 0.5         & 10847 $\pm$ 18      & 106 $\pm$ 35      & 187 $\pm$ 76      \\

  \hline
  \end{tabular}\\

    \end{center}
    \vskip 0.1 true cm \noindent Columns (1): the emission components; (2): RA and Dec of the components or spots; Columns (3)-(6) The Gaussian-fitted model parameters.
    \end{table*}
\begin{figure*}
   \centering

\includegraphics[width=18cm,height=21cm]{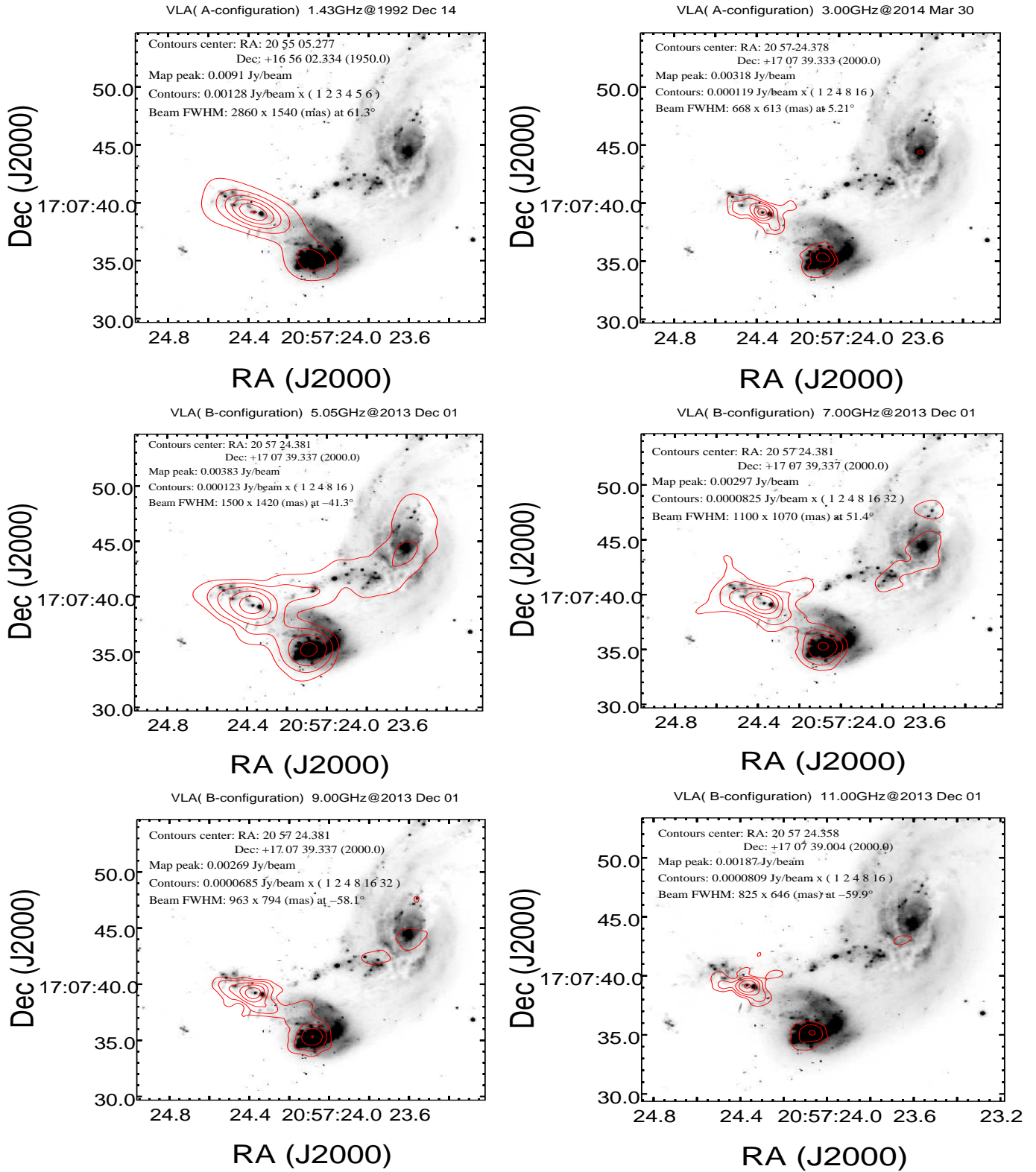}

    \caption{Multi-band radio contour maps of IIZw 096 from VLA archival data overlaid on the HST F814W(I) image}

    \label{L-Ka1}%
\end{figure*}
\addtocounter{figure}{-1}
\begin{figure*}
   \centering

\includegraphics[width=18cm,height=21cm]{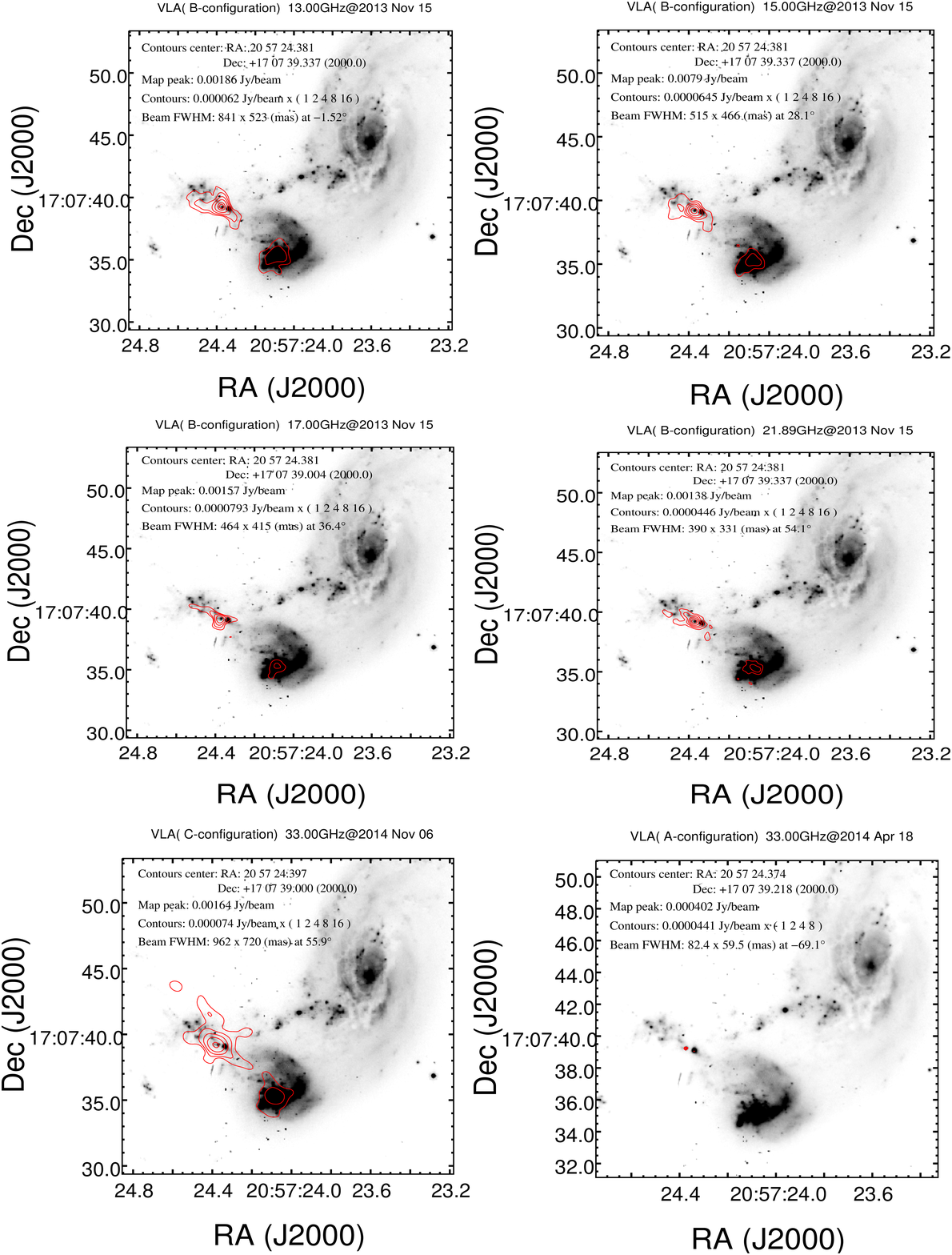}
    \caption{Continued}
    \label{L-Ka2}%
\end{figure*}
\begin{figure*}
   \centering
   \includegraphics[width=16cm,height=16cm]{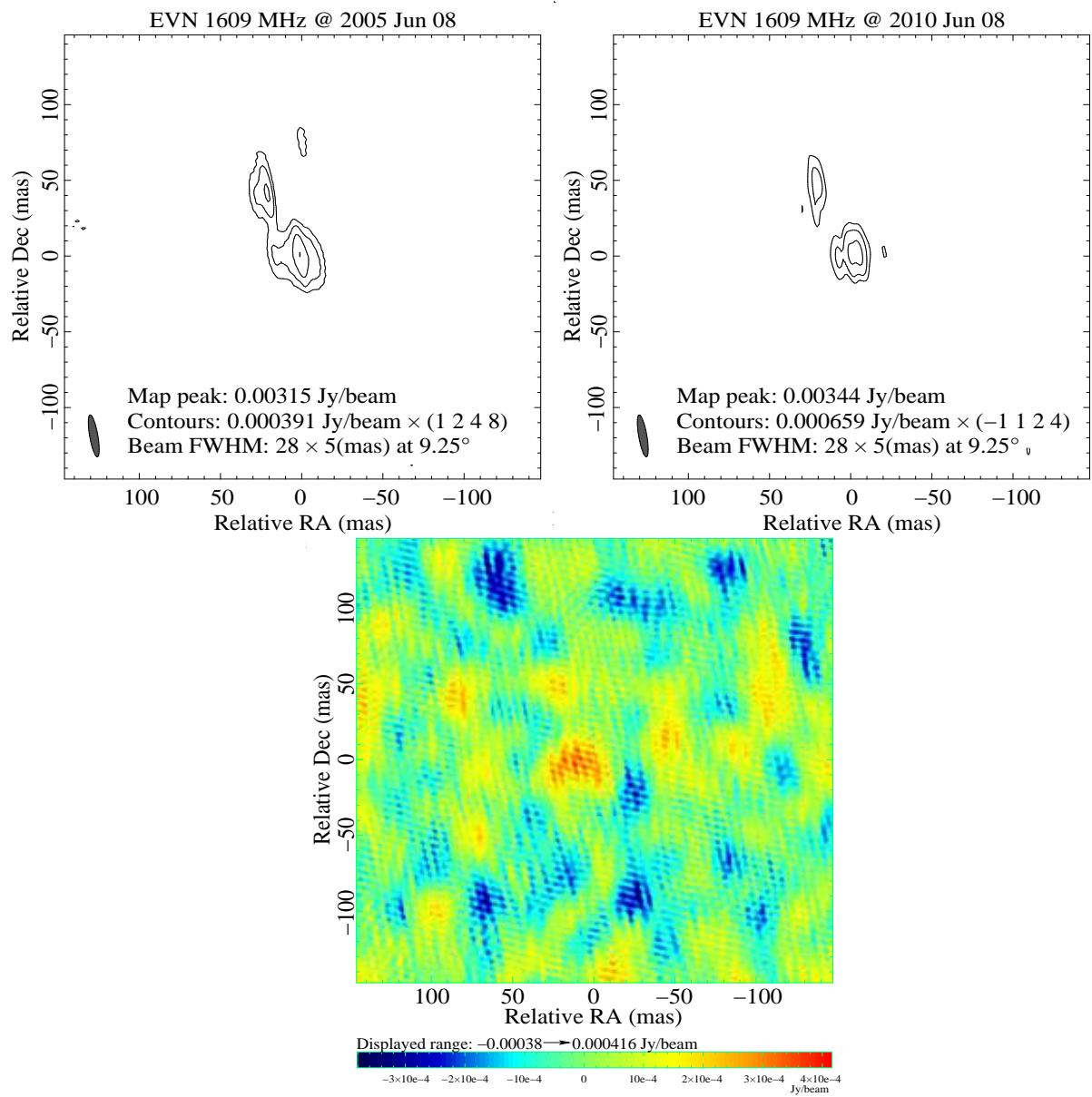}
      \caption{The high-resolution OH 1667/1665 MHz line emission from EVN observations \textbf{of IIZw 096}. Top panels are the images of two-epoch OH 1667 MHz line emission (V $\sim$ 10750-10950 \kms), which have been restored to the same beam size. The bottom panel is a dirty image of the OH-1665 line emission (V $\sim$ 11080-11320 \kms) obtained by combining the two-epoch EVN data.
      }
      \label{EVN2line}
\end{figure*}
\begin{figure*}
   \centering
\includegraphics[width=16cm,height=24cm]{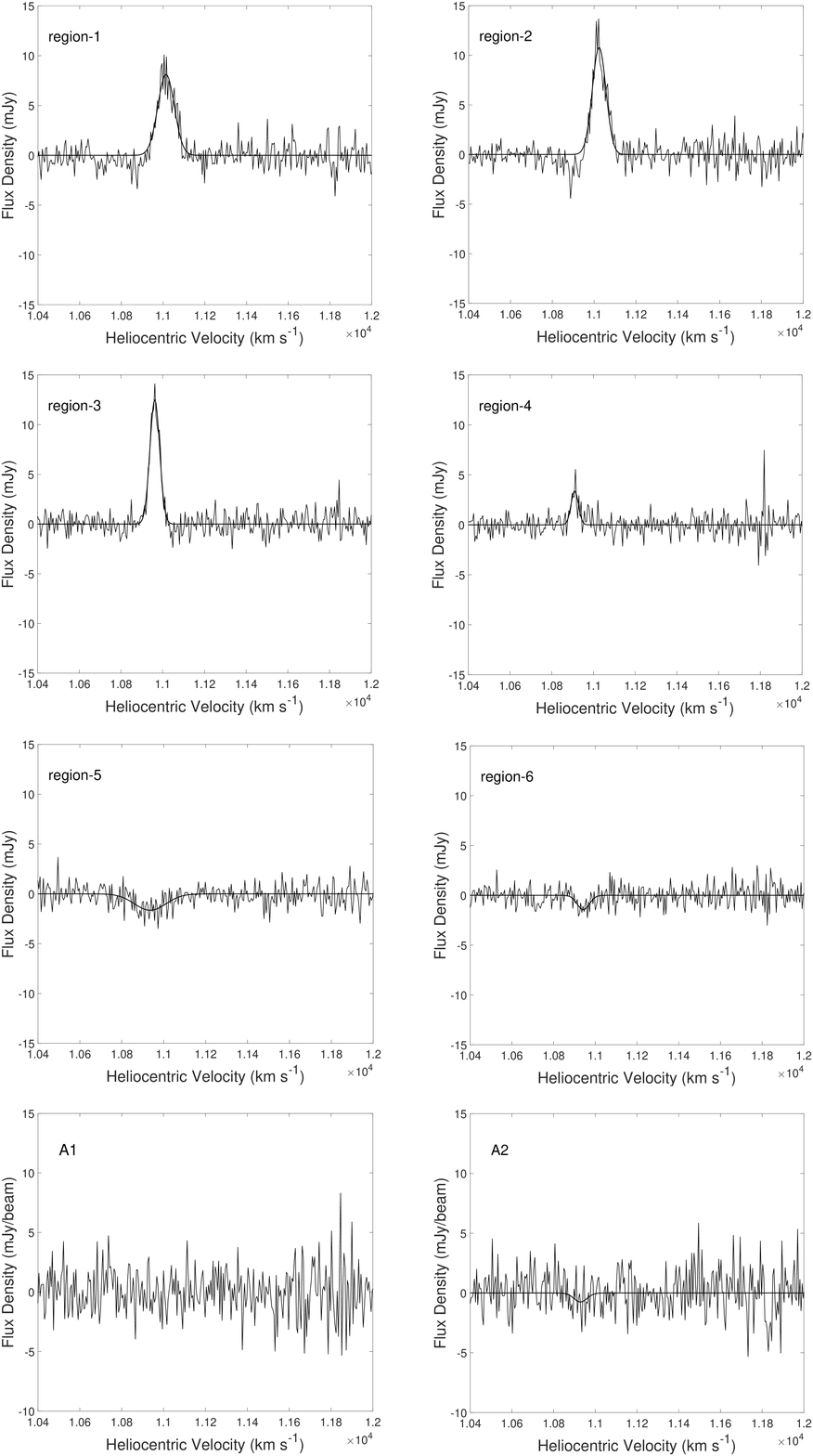}

      \caption{The extracted CO emission lines at various spots or regions of IIZw 096. The CO line profiles were fitted with one or two Gaussian components.  The blue and red lines are the fitted Gaussian components and the sum of these components, respectively}

    \label{COall}%
\end{figure*}
\addtocounter{figure}{-1}
\begin{figure*}
   \centering
\includegraphics[width=16cm,height=24cm]{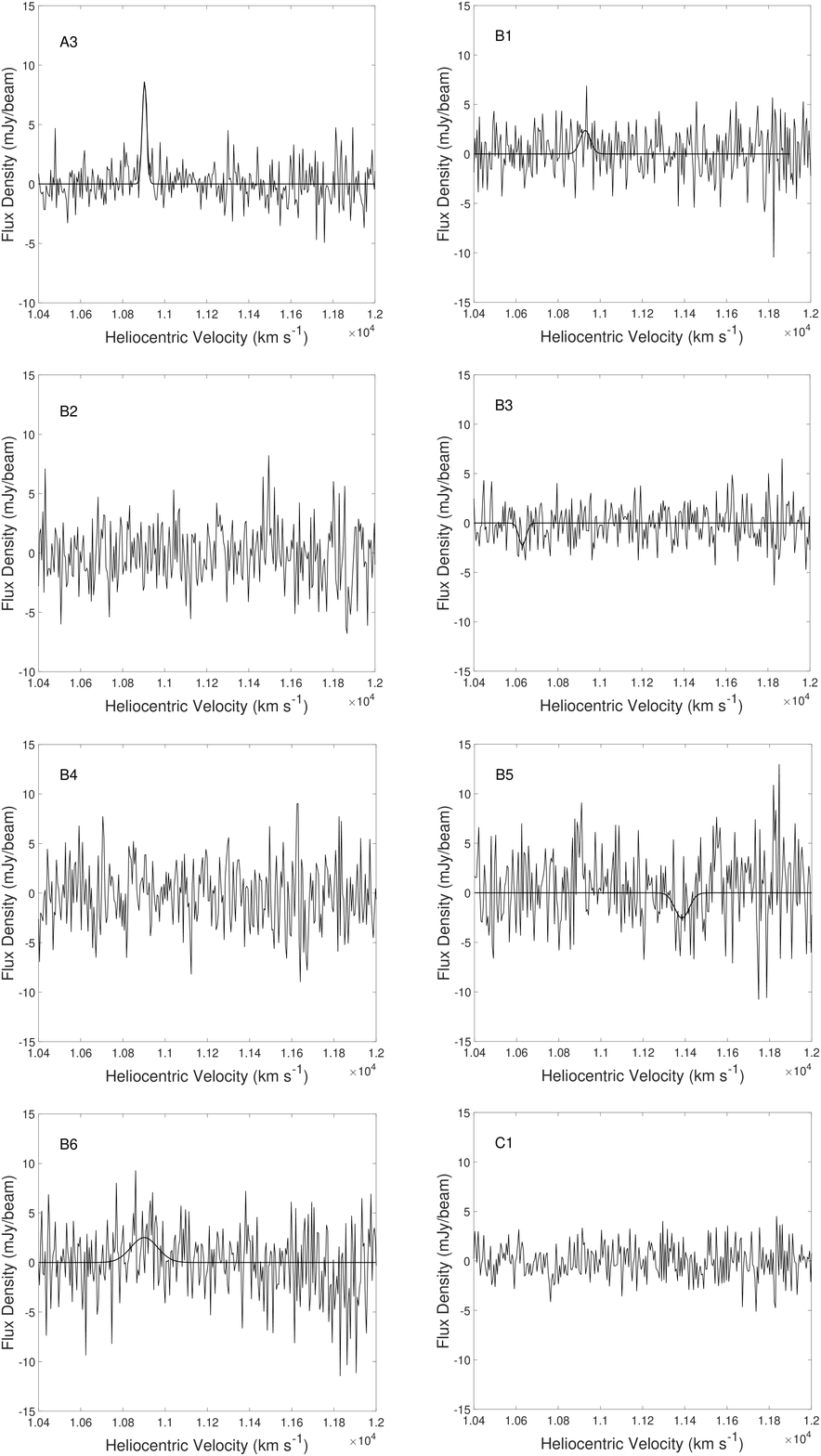}
\caption{Continued.}
 \label{COall}%
\end{figure*}
\addtocounter{figure}{-1}
\begin{figure*}
   \centering

\includegraphics[width=16cm,height=24cm]{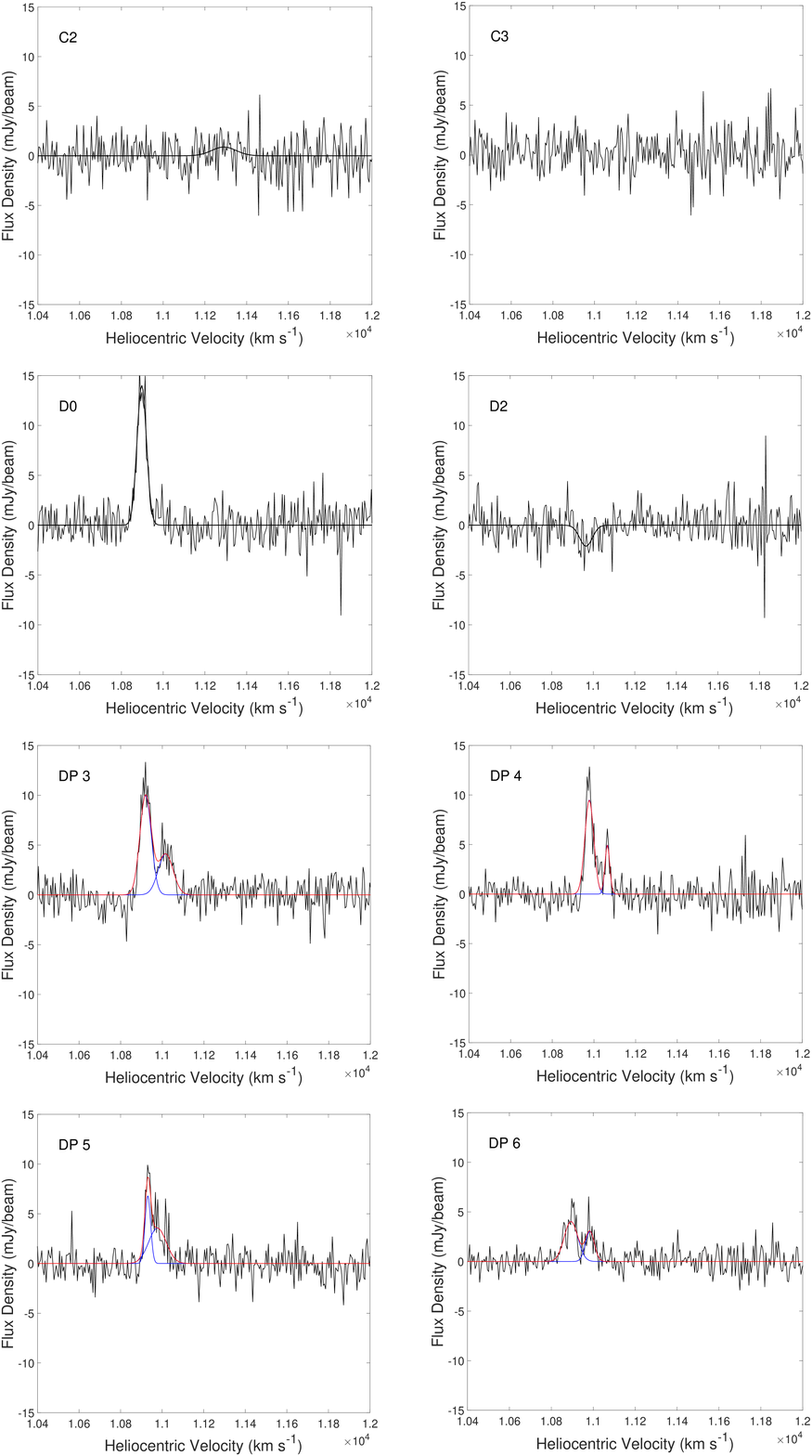}
      \caption{Continued.}

    \label{COall}%
\end{figure*}

\begin{figure*}
   \centering
 \includegraphics[width=16cm,height=24cm]{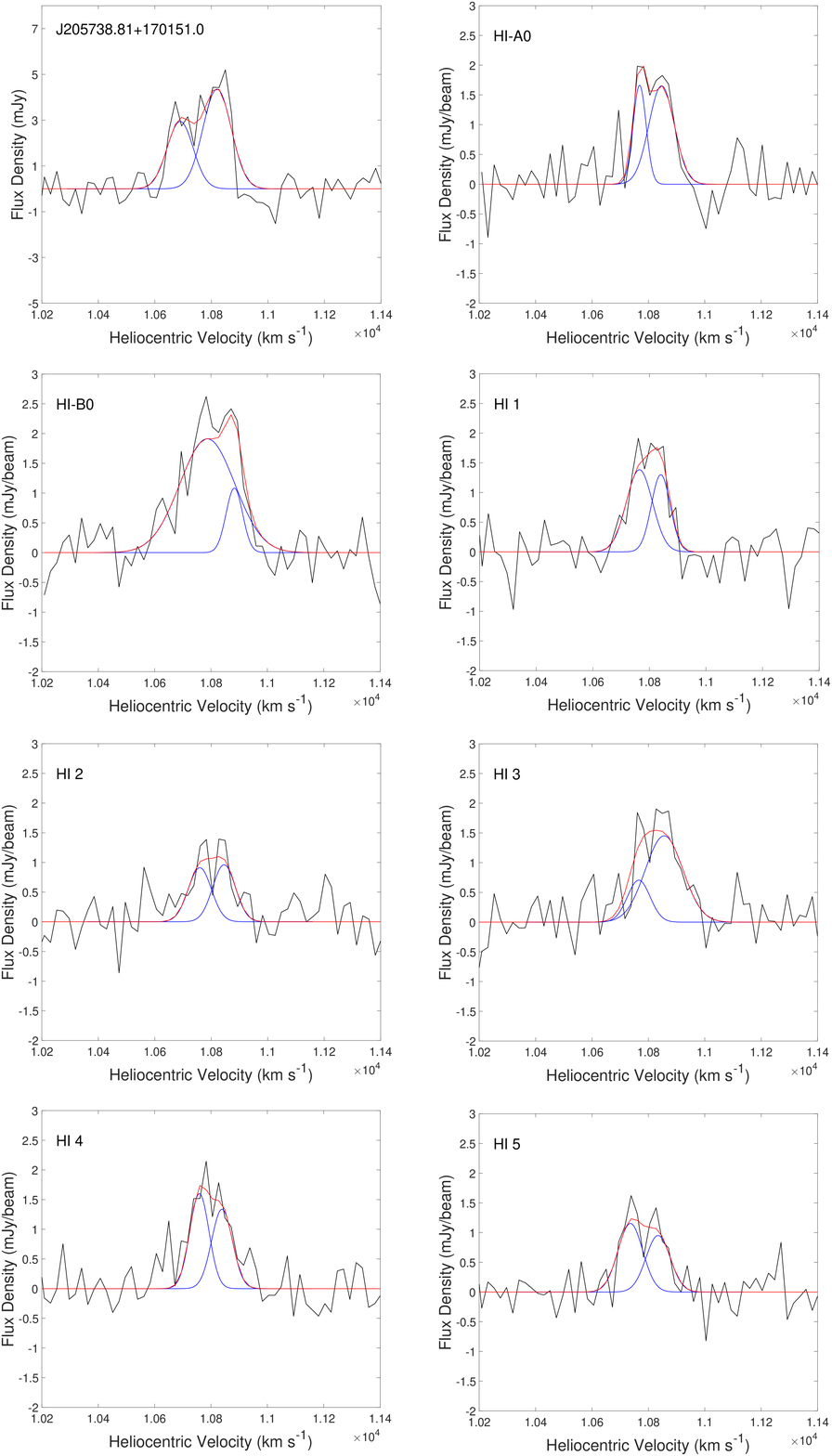}
        \caption{The extracted \HI emission lines at various spots or regions of IIZw 096. The blue and red lines are the fitted Gaussian components and the sum of these components, respectively}
      \label{HI-region1-7}
\end{figure*}
%---------------
\addtocounter{figure}{-1}
\begin{figure*}
   \centering
  \includegraphics[width=16cm,height=6cm]{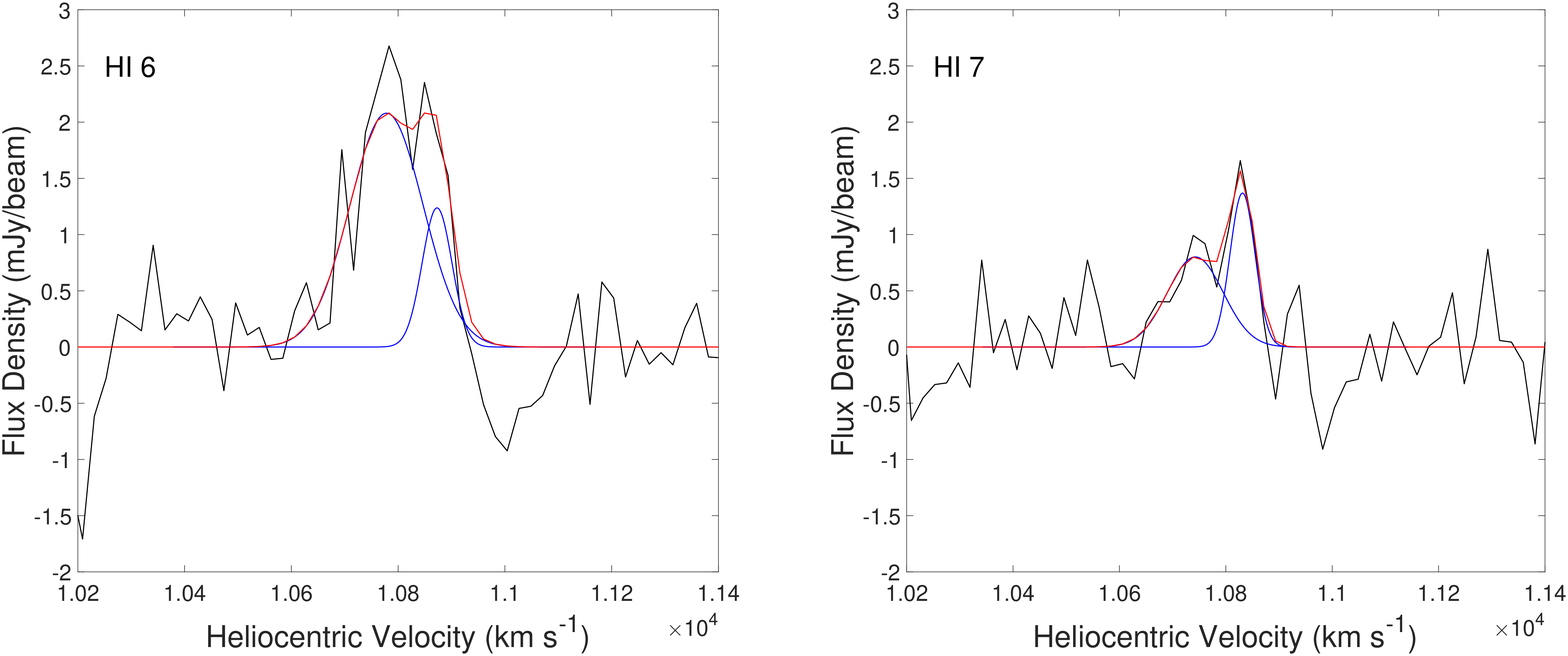}
        \caption{Continued.}
      \label{HI-region1-7}
\end{figure*}

%--------------------------------------------------------------------
\begin{figure*}
   \centering

\includegraphics[width=14cm,height=5.5cm]{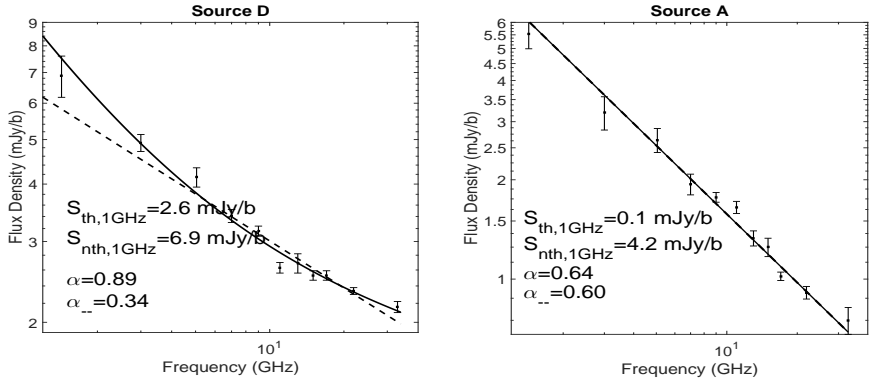}
      \caption{\textbf{Radio continuum spectral index by peaks of IIZw 096.} Radio continuum spectra of the peak flux densities of D and A from multi-band VLA projects listed in Table \ref{vladata}. The dashed and solid lines illustrate fitted results from the two equations, a$\times$$\nu$$^{-\alpha}$ and $S_{th} $$\times$ $\nu$$^{-0.1}$+$S_{nth}$ $\times$ $\nu$$^{-\alpha}$, respectively; where $S_{th} $ and $S_{nth}$ stands for non-thermal (synchrotron) and thermal (free-free) flux densities, respectively. }

    \label{ConstantDA-peak}
\end{figure*}

\newpage
%\section{Multi-bands of radio continue maps}

%----------------

\begin{figure*}
   \centering
   \includegraphics[width=16cm,height=7cm]{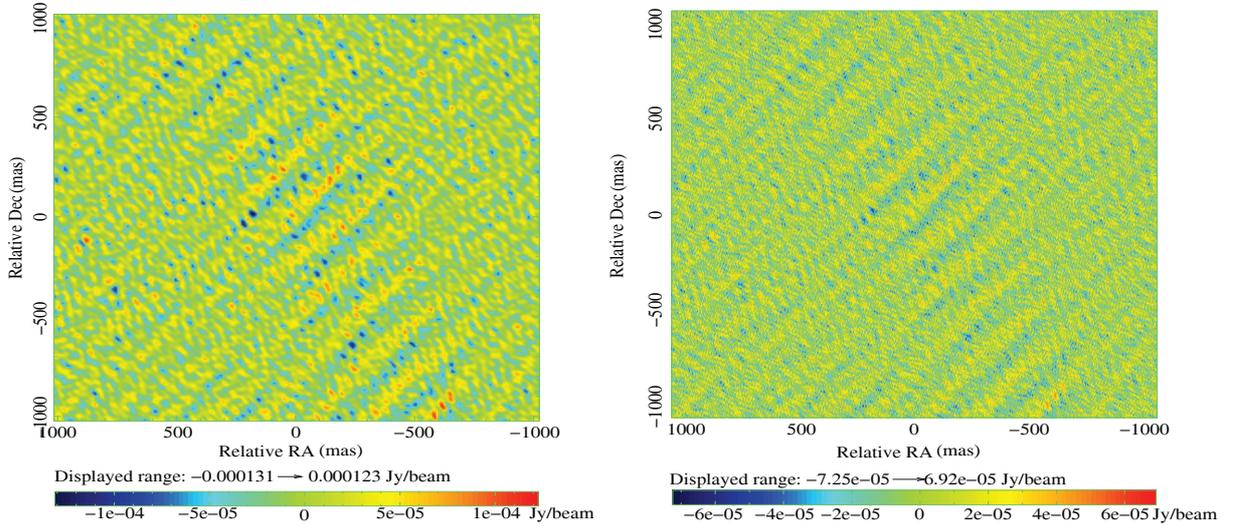}
      \caption{\textbf{The dirty maps of the continuum emission from the VLBA of IIZw 096.}Dirty maps of the continuum emission from the VLBA project BS0233 with different cell sizes, both centered  at RA: 20 57 24.377, Dec: +17 07 39.144. Right panel: The cell size is about 8 mas and the estimated dirty beam is about 22.28 mas $\times$ 36.8 mas at 38.0 degree,  1$\sigma$ noise is about 23.2 $\mu$Jy/beam. Left panel, Cell size is about 1 mas and the estimated dirty beam is about 4.83  $\times$ 11.5 at 2.1 degree, 1  $\sigma$ noise is about 14.8 $\mu$Jy/beam.
      }
      \label{vlbacontinuum}
\end{figure*}

%%----------------------------
\end{document}